\documentclass[useAMS,usenatbib]{mn2e}
\usepackage{graphicx}
\usepackage[usenames,dvipsnames]{xcolor}
\usepackage{url}
\usepackage{amssymb}
\usepackage{lscape}

\newcommand{\kms}{~km s$^{-1}$~}
\newcommand{\WR}{WR~48a~}
\newcommand{\WRE}{WR~48a}
\newcommand{\WC}{D2-3~}
\newcommand{\WCE}{D2-3}
\newcommand{\dotM}{~M$_{\odot}$~yr$^{-1}$~}
\newcommand{\XMM}{{\it XMM-Newton~}}
\newcommand{\XMME}{{\it XMM-Newton}}
\newcommand{\Chandra}{{\it Chandra~}}
\newcommand{\ChandraE}{{\it Chandra}}
\newcommand{\Swift}{{\it Swift~}}
\newcommand{\SwiftE}{{\it Swift}}
\newcommand{\Rosat}{{\it ROSAT~}}
\newcommand{\RosatE}{{\it ROSAT}}
\newcommand{\apj}{ApJ}
\newcommand{\xspec}{{\sc xspec~}}
\newcommand{\xspecE}{{\sc xspec}}

\def\kms{\mbox{~km\,s$^{-1}$\/}}

\def\utw{\smash{\rlap{\lower5pt\hbox{$\sim$}}}}
\def\udtw{\smash{\rlap{\lower6pt\hbox{$\approx$}}}}

\def\farcs{\hbox{$.\!\!^{\prime\prime}$}}

\title[A multi-wavelength view on \WR]
{A multi-wavelength view on the dusty Wolf-Rayet star
\WR\thanks{Based on observations made with the Southern African Large
Telescope (SALT) %}
under program 2012\_1\_POL\_OTH\_1 (PI: Toma Tomov).}
 }
\author[S.A.Zhekov et al.]{Svetozar A. Zhekov$^1$
\thanks{E-mail: szhekov@space.bas.bg 
}, 
Toma Tomov$^2$, Marcin P. Gawronski$^2$, Leonid N.
Georgiev$^3$\thanks{deceased,
2012 Dec 26}, 
\newauthor Jura Borissova$^{4,5}$, 
Radostin Kurtev$^{4,5}$, 
Marc Gagn\'{e}$^6$ and Marcin Hajduk$^7$ \\
 $^1$Space Research and Technology Institute, Akad. G.
Bonchev str., bl.1, Sofia 1113, Bulgaria\\
$^2$Centre for Astronomy, Faculty of Physics, Astronomy and
Informatics, Nicolaus Copernicus University, Grudziadzka 5, 87-100
Torun, Poland \\
$^3$Instituto de Astronomía, Universidad Nacional Autónoma de 
M\'{e}xico, Apartado Postal 70-264, CP 04510 M\'{e}xico DF, Mexico\\
$^4$ Instituto de F\'{i}sica y Astronom\'{i}a, Facultad de Ciencias, 
Universidad de Valpara\'iso, Av. Gran Breta\~na 1111, Playa Ancha, 
Casilla 5030, \\ Valpara\'{i}so, Chile \\
$^5$ Millennium Institute of Astrophysics, MAS, Santiago, Chile\\
$^6$Department of Geology and Astronomy, West Chester
University, West Chester, PA 19383, USA\\
$^7$Nicolaus Copernicus Astronomical Center, ul. Rabianska 8, 87-100,
Torun, Poland
}

\date{}

\pagerange{\pageref{firstpage}--\pageref{lastpage}} \pubyear{2002}

\begin{document}

\maketitle

\label{firstpage}

\begin{abstract}
We present results from the first attempts to derive various physical
characteristics of the dusty Wolf-Rayet star \WR based on  a 
multi-wavelength view of its observational properties. This is done on 
the basis of new optical and near-infrared spectral observations and on 
data from various archives in the optical, radio and X-rays. 
The optical spectrum of \WR is acceptably well represented by
a sum of two spectra: of a WR star of the WC8 type and of a WR star of
the WN8h type.
The strength of the interstellar absorption features in the 
optical spectra of \WR and the near-by stars \WC and D2-7 (both
members of the open cluster Danks 2) indicates that \WR is located 
at a distance of $\sim4$ kpc from us.
\WR is very likely a thermal radio source and for such a 
case and smooth (no clumps) wind its radio emission suggests a 
relatively high mass-loss rate of this dusty WR star
($\dot{M} \approx \mbox{a few} \times 10^{-4}$\dotM).
Long timescale (years) variability of \WR is established in the
optical, radio and X-rays. 
Colliding stellar winds likely play a very important role in the 
physics of this object. However, some LBV-like (luminous blue 
variable) activity could not be excluded as well.

\end{abstract}


\section{Introduction}
\WR is a carbon-rich (WC) Wolf-Rayet star (WR) that was discovered in
a near-infrared survey by Danks et al. (1983). It is located
inside the G305 star-forming region in the Scutum Crux arm of the
Galaxy. Two compact infrared clusters (Danks 1 and 2) are found in its
vicinity (within $2'$) which indicates that this WR star likely 
originates from one or the other \citep{danks_84}.
The most pronounced characteristics of \WR deduced from observations
are: 
(a) its optical extinction is very high, A$_V = 9.2$~mag
\citep{danks_83}
(b) it is a strong and variable infrared source (e.g.,
\citealt{williams_12});
(c) it is very luminous and variable in X-rays 
(\citealt{zhgsk_11}, 2014).

The infrared variability of \WR suggests that it is a long-period
episodic dust maker with short-term `flares' superimposed on a much
gradually changing emission (\citealt{williams_95},
\citealt{williams_03}).
A recent study revealed a recurrent dust formation 
on a timescale of more than 32 years which also indicates that \WR 
is very likely a wide colliding-wind binary \citep{williams_12}.
However, the case of \WR might not be that simple.
For example, \citet{hindson_12} reported detection of a 
{\it thermal} radio source (spectral index 
$\alpha = 0.6$, F$_{\nu} \propto \nu^{\alpha}$) 
associated with \WRE, while the wide
colliding-wind binaries are in general {\it non-thermal} radio sources 
\citep{do_00}.

Nonetheless, we note that the X-ray properties of \WR provide
additional support for the binary nature of this WC star. 
Analysis of the \XMM and \Chandra spectra of \WR showed that
(\citealt{zhgsk_11}, 2014):
(i) its X-ray emission is of thermal origin;
(ii) this is the most X-ray luminous WR star in the Galaxy detected so
far, after the black-hole candidate Cyg X-3;
(iii) the X-ray emission is variable and the same is valid for the
X-ray absorption to this object.
All these X-ray characteristics are well explained in the framework of
the colliding-stellar-wind (CSW) picture although specific
information from different spectral domains is yet needed for carrying
out a detailed quantitative modelling.

In this study, our goal was to undertake the first attempts to
derive various physical characteristics of \WR in different spectral
domains.  To do so, we obtained new optical and near-infrared spectra
of \WRE.  Also, we made use of some data from the 
optical, radio and X-ray archives of various ground-based and space 
observatories. All this, the new and archival data, allowed us to
derive valuable pieces of information on the evolution of the physical
characteristics of \WRE.  We could thus start building a global 
picture of this fascinating object based on a multi-wavelength view of 
its properties.
In Section \ref{sec:observations}, we review the observational
information in different spectral domains. In Section 
\ref{sec:results}, we describe our results. We discuss the results
from our analysis in Section \ref{sec:discussion} and list our 
conclusions in Section \ref{sec:conclusions}.

\section{Observations and data reduction}
\label{sec:observations}

\begin{figure}
\begin{center}
\centering\includegraphics[width=\columnwidth]{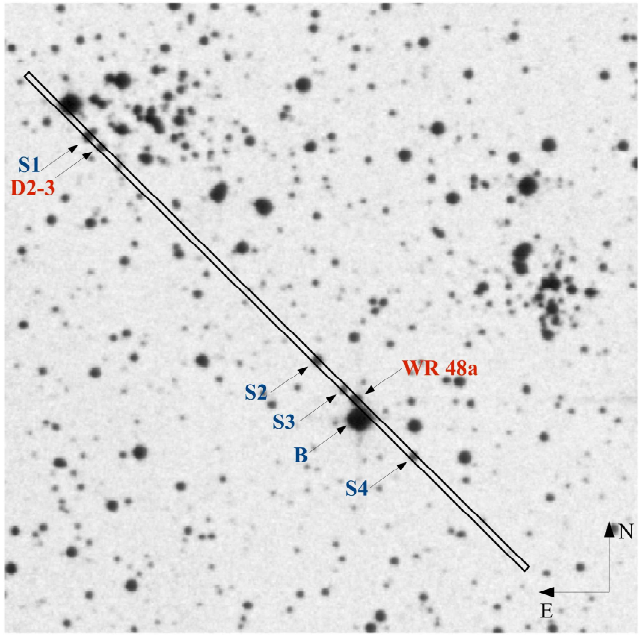}
\end{center}
\caption{A $5\times5$ arcmin field in the vicinity of \WR from the 
MAMA.R.ESO image.  The rectangle represents the RSS slit 
position on the sky. The objects, whose spectra were used in this
study, are marked by arrows: 
\WRE, \WC and several additional stars 
B, S1, S2, S3 and S4 (see the text for details).
}
\label{fig:salt_slit}
\end{figure}

\subsection{Optical and NIR}
\label{subsec:optical}
We obtained new optical and near-infrared (NIR) spectra of \WR with
the Southern African Large Telescope (SALT; \citealt{buck_06};
\citealt{odon_06}), the 1.9m telescope at the South 
African Astronomical Observatory and the New Technology
Telescope (NTT) at the European Southern Observatory (ESO).
Our search in the archives of various ground-based observatories in
the southern hemisphere found one useful optical observation with the
Anglo-Australian Telescope (AAT) at the Anglo-Australian Observatory.
We give next some basic information on these data and the corresponding 
data reduction.

\begin{figure*}
\centering\includegraphics[width=6in,height=3.44in,clip=true]{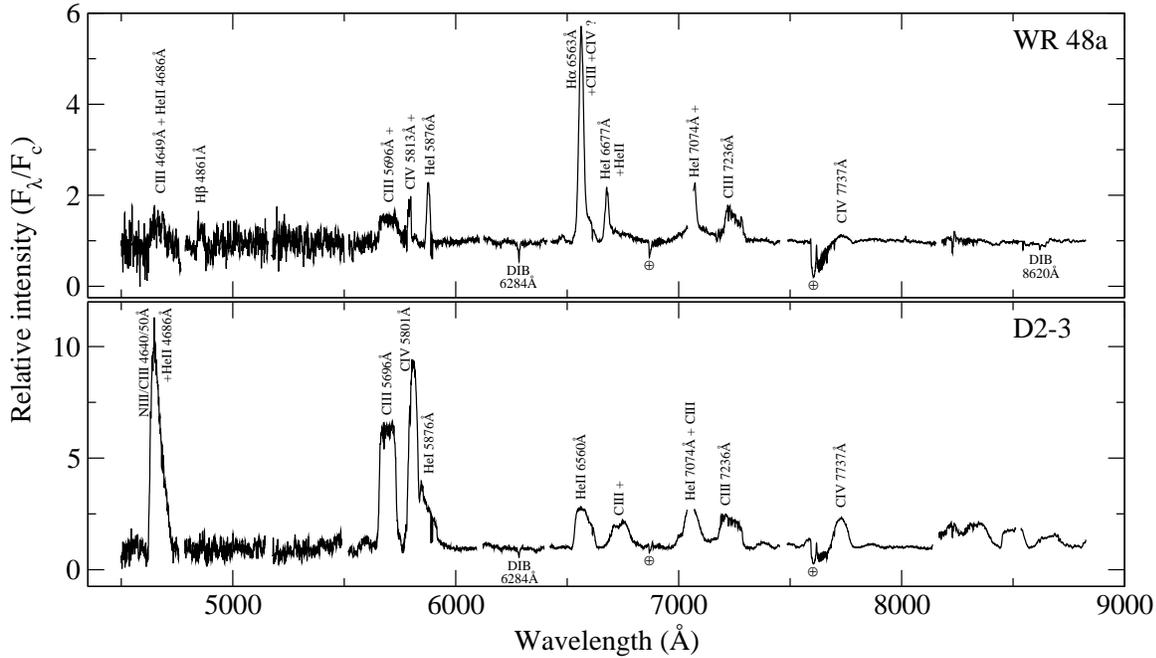}
\caption{
Normalized to the local continuum SALT spectra of \WR
(upper panel) and \WC (lower panel). The gaps in the spectra reflect
the RSS inter-chip gaps. The line identification is based on
\citet{williams_12} and on \citet{mauer_09} for \WR and \WCE,
respectively. The strongest atmospheric absorption
bands are marked with Earth symbols.
}
\label{fig:salt_spec}
\end{figure*}

\begin{figure*}
\centering\includegraphics[width=6in,height=2.09in,clip=true]{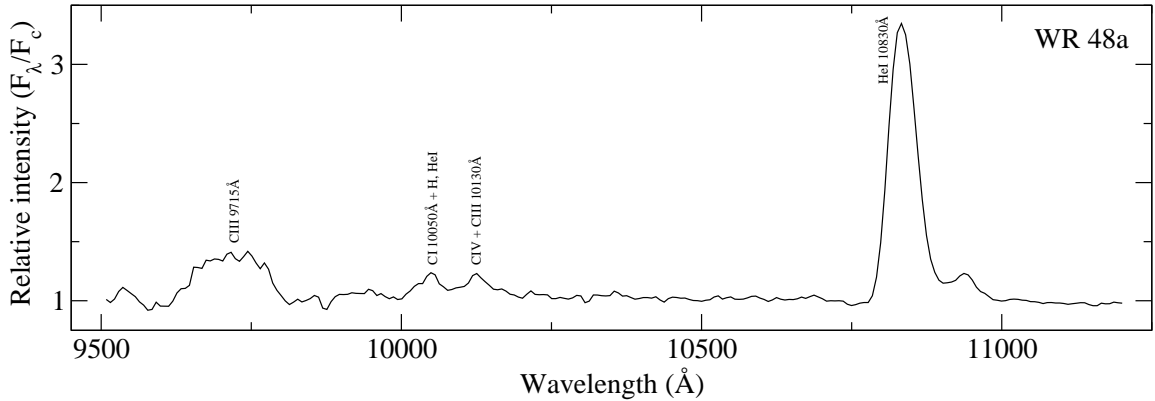}
\caption{
The NTT SofI spectrum of \WR normalized to the
local continuum.
The line identification is based on \citet{williams_12}.
}
\label{fig:ntt_spec}
\end{figure*}

{\it SALT.}
The SALT spectra of \WR were obtained with the Robert
Stobie Spectrograph (RSS; \citealt{burgh_03}; \citealt{kobul_03})
on 2012 May 27. The spectrum of a bright, nearby B star (CPD-62 3058,
see Fig.~\ref{fig:salt_slit}) was obtained with the same setup on 2012 
May 24. In the RSS long-slit spectroscopy mode, the volume phase
holographic (VPH) gratings PG1800 and PG2300 were used in the spectral 
range 4000-8800\,\AA. The whole spectrum in this region was obtained
in five shots covering the subregions 4050-5100\,\AA, 4880-5850\,\AA,
5810-6670\,\AA, 6640-7840\,\AA\ and 7790-8840\,\AA, respectively.
Spectra of ThAr, Ne and Xe comparison arcs were obtained to calibrate
the wavelength scale. The slit width was 1\farcs5 and its position on
the sky is shown in Fig.~\ref{fig:salt_slit}. The position of the slit
was chosen in this way, to be able to obtain the spectra of \WR and
the WC8 star \WC (2MASS J13125770-6240599) simultaneously. As one can 
see in Fig.~\ref{fig:salt_slit}, four additional stars are well located 
in the slit and their spectra were also considered in this paper (see 
Appendix~\ref{append:stars}). The spectral resolution in the particular 
subregions, estimated by the  full widths at the half maximum (FWHM) of 
the extracted comparison spectra lines, is as follows: 
$1.52\pm0.05$\,\AA, $1.40\pm0.17$\,\AA, $1.20\pm0.05$\,\AA, 
$1.66\pm0.07$\,\AA ~and $1.36\pm0.05$\,\AA. For the flux 
calibration, we used spectra of the spectrophotometric standard stars 
LTT\,4364, LTT\,7987 and Feige\,110, obtained in the nights of May 
24 and 27.

The initial data reduction, including bias and overscan 
subtraction, gain and cross-talk corrections, trimming and mosaicking,
was done with the SALT science pipeline \citep{craw_10}. We
used the {\sc dcr} software written by W.~Pych (see 
\citealt{pych_04} for details) to remove the cosmic rays from the
observations. To perform the flat field corrections, the wavelength 
calibration and to correct the distortion and the tilt of the frames
we used the standard {\sc iraf}\footnote{{\sc iraf} is 
distributed by the National Optical Astronomy Observatories, which are 
operated by the Association of Universities for Research in Astronomy, 
Inc., under cooperative agreement with the National Science 
Foundation.} tasks in the {\sc twospec} package. The nights of our 
observations were not photometric. Moreover, SALT is a telescope with 
a variable pupil which makes the absolute flux calibration impossible. 
Thus, we calibrated the spectra in relative flux units using an 
average sensitivity curve. This allowed us to derive  the relative 
spectral energy distribution of the studied objects. The normalized 
SALT spectra of \WR and \WC are presented in Fig.~\ref{fig:salt_spec}.

A low-resolution spectrum of \WR was acquired on 2013 July 17 with the 
Grating Spectrograph with a SITe CCD mounted on the 1.9\,m Radcliffe 
telescope at the South African Astronomical Observatory (SAAO).  
Grating number\,7 with 300 lines\,mm$^{-1}$ and a~slit width of 
$1\farcs5$ were used. 
To calibrate the spectra in relative flux, the spectrophotometric 
standard star EG\,21 was used.  All the data reduction and 
calibrations were carried out with the standard {\sc iraf} 
procedures. The extracted and flux-calibrated spectrum covers the 
range $\sim$4200--7500\,{\AA} with a resolution $5.46\pm0.69$\,\AA, 
derived from the FWHM of the comparison spectrum lines.

{\it ESO}.
\WR was observed on 2011 Apr 14 (total integration time of 180 sec)
using the infrared spectrograph and imaging camera SofI
\citep{moor_98} in long-slit
mode on NTT at La Silla Observatory (ESO). The J-spectrum scale is
6.96 °A/pix that is its resolving power is R $\sim1000$. The spectral
resolution is 13.92 °A (the full width at the half maximum). Bright 
stars of spectral type G were observed as a measure of the atmospheric
absorption. The spectra were reduced using the standard {\sc iraf}
procedures: correction for the bad pixels, bias level and sky emission
lines subtraction, flat fielding, spectrum extraction, wavelength
calibration and  telluric correction (for more details see
\citealt{chene_12}). Unfortunately, no flux calibration was possible 
and this is why the J-spectrum was normalized to the stellar continuum 
(Fig.~\ref{fig:ntt_spec}) and used for spectral line analysis only.

{\it AAT.}
We downloaded from the AAT archive two
spectra of \WR (the same ones used by \citealt{williams_12}) obtained 
with the RGO spectrograph on  1993 June 21. The spectra were reduced 
and calibrated by the use of the standard {\sc iraf} packages. The
spectrophotometric standard star LTT\,377, observed during the same
night, was used for the relative flux-calibration of the spectra. The 
blue spectrum covers the region from $\sim3500$\,\AA\ to 
$\sim6700$\,\AA\ with a resolution $9.65\pm0.59$\,\AA\ estimated on 
the base of the comparison spectrum lines FWHM. The red spectrum 
covers the region from $\sim5300$\,\AA\ to $\sim11000$\,\AA\ and its 
resolution is $21.5\pm1.5$\,\AA.

The equivalent widths (EW) and the FWHM of the strongest emission
lines, interstellar absorption lines and diffuse interstellar
bands (DIBs), measured in all our spectra are provided in Tables~1, 
\ref{tab:wc8_lines} and \ref{tab:dibs}. The measurements were done
with the {\sc iraf} tasks {\sc splot} and {\sc spectool} 
fitting gaussians to the lines. The error estimates were computed 
directly in {\sc splot} and {\sc spectool} by running a number 
of Monte Carlo simulations based on preset instrumental parameters.

\begin{figure}
\centering\includegraphics[width=\columnwidth]{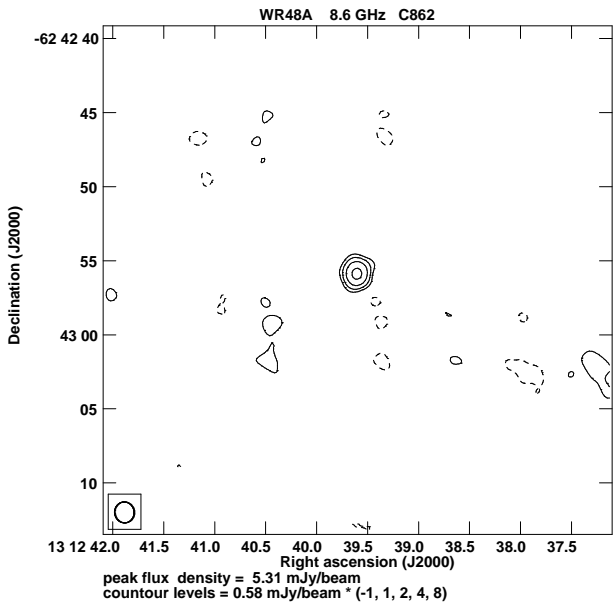}
\caption{
Radio map of \WR at 4.8\,GHz based on data from the ATCA project 
C862. The first contour in map represents $3\sigma$ detection limit.
}
\label{fig:atca}
\end{figure}

\begin{figure*}
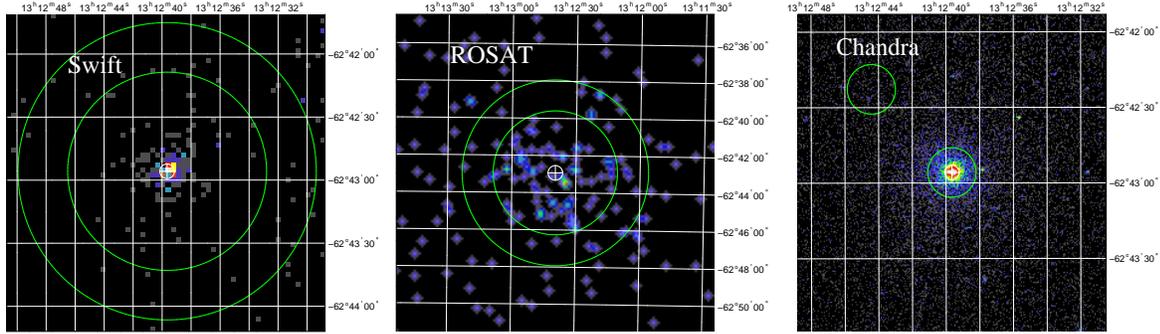

\begin{center}
\includegraphics[width=2.in, height=1.78in]{fig5a.eps}
\includegraphics[width=2.in, height=1.78in]{fig5b.eps}
\includegraphics[width=2.in, height=1.78in]{fig5c.eps}
\end{center}
\caption{
Raw \Swift (ObsId 00031900028), \Rosat (ObsId: rp190249n00) and
\Chandra ACIS-I (ObsId: 8922) images.
R.A. (J2000) and decl. (J200)
are on the horizontal and vertical axes, respectively. The cirlced
plus sign denotes the optical position of \WR (SIMBAD). The inner
circle marks the source extraction region. For the \Swift and \RosatE,
the annulus denotes the background extraction region while the
background region is another circle (upper left from the image centre)
for the case of \Chandra ACIS-I.
}
\label{fig:xrayimages}
\end{figure*}

\begin{figure*}
 \includegraphics[width=6.65in, height=8.0in,angle=-90]{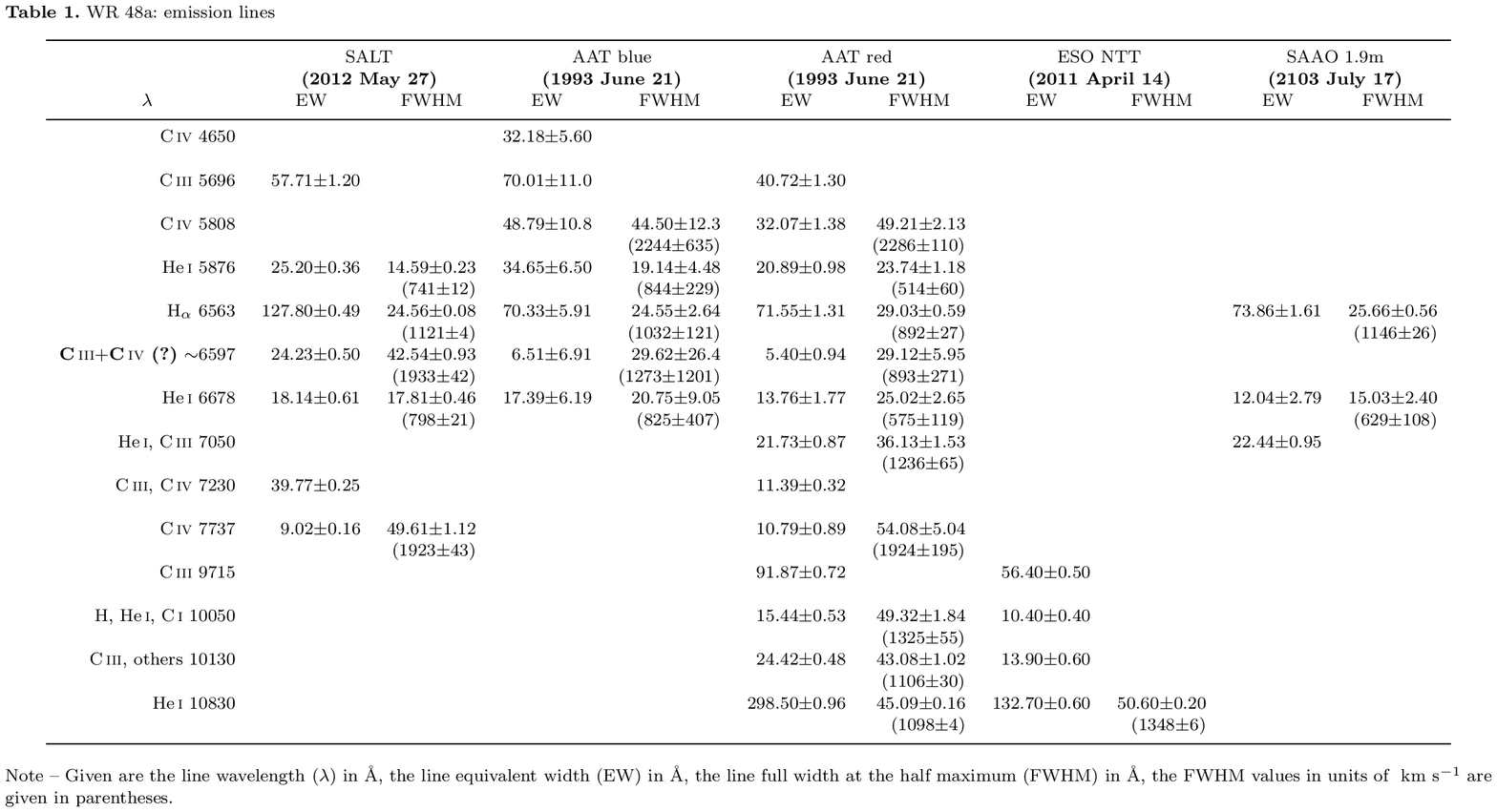}
\end{figure*}

\subsection{Radio}
\label{subsec:radio}
\WR was observed in numerous occasions with the Australia Telescope
Compact Array (ATCA) operated by the Australia Telescope National
Facility. We have checked the available data in the Australia
Telescope Online Archive\footnote{\url{http://atoa.atnf.csiro.au/}} 
and made use of the radio observations taken 
on 2000 Feb 27 (project code C862; ATCA configuration 6A) and 
on 2006 Dec 8 (project code C1610; ATCA configuration 6B). 
During the project C862, \WR was observed at four frequencies 1.4, 
2.5, 4.8 \& 8.6\,GHz with a bandwidth of 128\,MHz. However due to 
data quality, we were able to recover fluxes only at 4.8 \& 8.6\,GHz. 
At the two lower frequencies, there were a lot of interferences in the
data that resulted in unreliable maps. Because of that, we were unable
to make any conclusions about the \WR radio fluxes at 1.4 and 2.5 GHz.
In the
project C1610, \WR was observed at two frequencies 4.8 \& 8.6\,GHz with 
a bandwidth of 128\,MHz (Fig.~\ref{fig:atca}). We were able to reduce 
all data and estimate the \WR 
fluxes at both used radio bands. 1934-638 was used as a primary flux 
calibrator in both bands but different phase calibrators were chosen:
1251-713 for C862 and 1352-63 for C1610, respectively. The whole 
data reduction process was carried out using the standard NRAO 
{\sc aips}\footnote{\url{www.aips.nrao.edu/index.shtml}} procedures. 
The task {\sc imagr} was used to produce the final total intensity 
images of \WRE. Radio fluxes were then measured, by fitting Gaussian 
model using the {\sc aips} task {\sc jmfit}.

\subsection{X-rays}
\label{subsec:xrays}
To broaden the study of the X-ray properties of \WRE, especially, that
on its long-term X-ray variability, we have checked the archives of
the modern X-ray observatories (e.g., \ChandraE, \XMME, \RosatE, 
\SwiftE). Thus, in addition to the already analysed and published data
(see \citealt{zhgsk_11}, 2014), we found two \Rosat 
and 33 pointed \Swift observations. We also gave some consideration 
to the \Chandra ACIS-I data on \WR which are heavily piled up and thus
cannot be used for spectral analysis.

\SwiftE.
\WR was observed multiple times between 2010 Dec 15 and 2013 Feb 27 
with typical exposure time of 2,400 - 5,000 sec.
From all the 33 pointed observations (ObsID from 00031900001 to
00031900033), one (ObsID 00031900017) was not included in this
analysis due to its very short exposure time ($\sim 70$~sec). 
Following the Swift XRT Data Reduction
Guide\footnote{\url{http://swift.gsfc.nasa.gov/analysis/xrt_swguide_v1_2.pdf}},
we extracted the source and background spectra for each observation.
Extraction regions had the same shape and size for each data set 
(see Fig.~\ref{fig:xrayimages}).
For our analysis, we used the response matrix 
(swxpc0to12s6\_20010101v014.rmf)  provided by
the most recent (2014 Feb 02) \Swift calibration
files\footnote{\url{http://heasarc.gsfc.nasa.gov/docs/heasarc/caldb/swift/}}
and we also used the package {\it xrtmkarf} to construct the ancillary 
response file for each data set.

\RosatE.
\WR fell in the \Rosat field of view in two occasions in 1997 February. 
The corresponding PSPC data sets are
rp190020n00 (Feb 5; 20.5 arcmin off axis; exposure time 740 sec) and
rp190249n00 (Feb 23; 34.3 arcmin off axis; exposure time of 1080 sec).
Following the recommendations for the \Rosat Data
Processing\footnote{\url{http:
//heasarc.gsfc.nasa.gov/docs/rosat/rhp_proc_analysis.html}}, we
extracted the source and background spectra
(see Fig.~\ref{fig:xrayimages}). Since the data were taken after 1991
Oct 14, we adopted the response matrix pspcb\_gain2\_256.rmf and we
used the package {\it pcarf} to construct the ancillary response file
for each observation. Due to the limited photon statistics of the
\Rosat spectra of \WR (each has no more than 50 source counts), we 
combined them and used the resultant spectrum in our analysis (see
Section~\ref{subsec:xray_lc}).

\ChandraE.
\WR was detected by \Chandra (ObsID: 8922) but the pileup in the 
ACIS CCD is very high (see footnote 5 in \citealt{zhgsk_11}).
Nevertheless, we used the Chandra Interactive Analysis of
Observations 4.4.1\footnote{For details, see
\url{http://cxc.harvard.edu/ciao/}}
data analysis software to formally extract an X-ray 
source spectrum and the background emission in the source vicinity 
(see Fig.~\ref{fig:xrayimages}). 
We 
could thus estimate the source background-subtracted count rate. 
Despite being affected by the strong pileup, we made some use of this
parameter in the discussion of the X-ray light curve of \WR 
(see Section~\ref{subsec:xray_lc}).

For the spectral simulations in this study, we made use of
version 11.3.2 of \xspec \citep{Arnaud96}.

%
%  Table 1 was here
%

\section{Results}
\label{sec:results}

\subsection{Spectral lines}
\label{subsec:lines}
For \WR and \WCE, we measured the equivalent width (EW) and the full 
width at the half maximum (FWHM) for all the spectral lines that were 
detected in the optical and NIR spectra at hand: SALT, AAT blue and 
red channels, ESO NTT, SAAO 1.9m telescopes. 
Before estimating the FWHM values in units of
\kms, we corrected them for the instrumental broadening:
$FWHM_0 = \sqrt{FWHM^2 - FWHM_{sp.res.}^2}$ ($FWHM_{sp.res.}$ is the
spectral resolution for the given spectrum, see
Section~\ref{subsec:optical}).  

\begin{figure*}
\begin{center}
\includegraphics[width=3.in, height=1.7in]{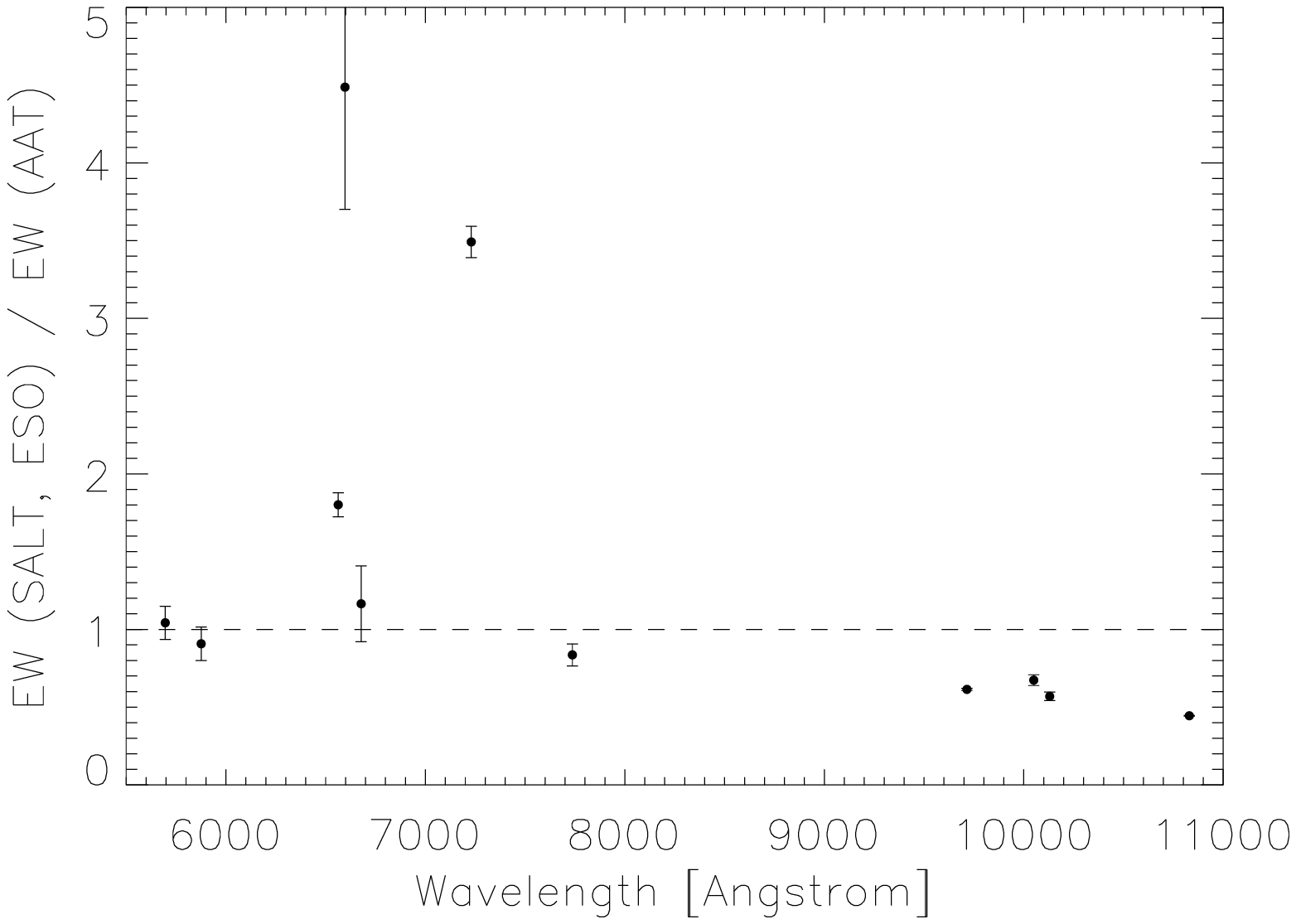}
\includegraphics[width=3.in, height=1.7in]{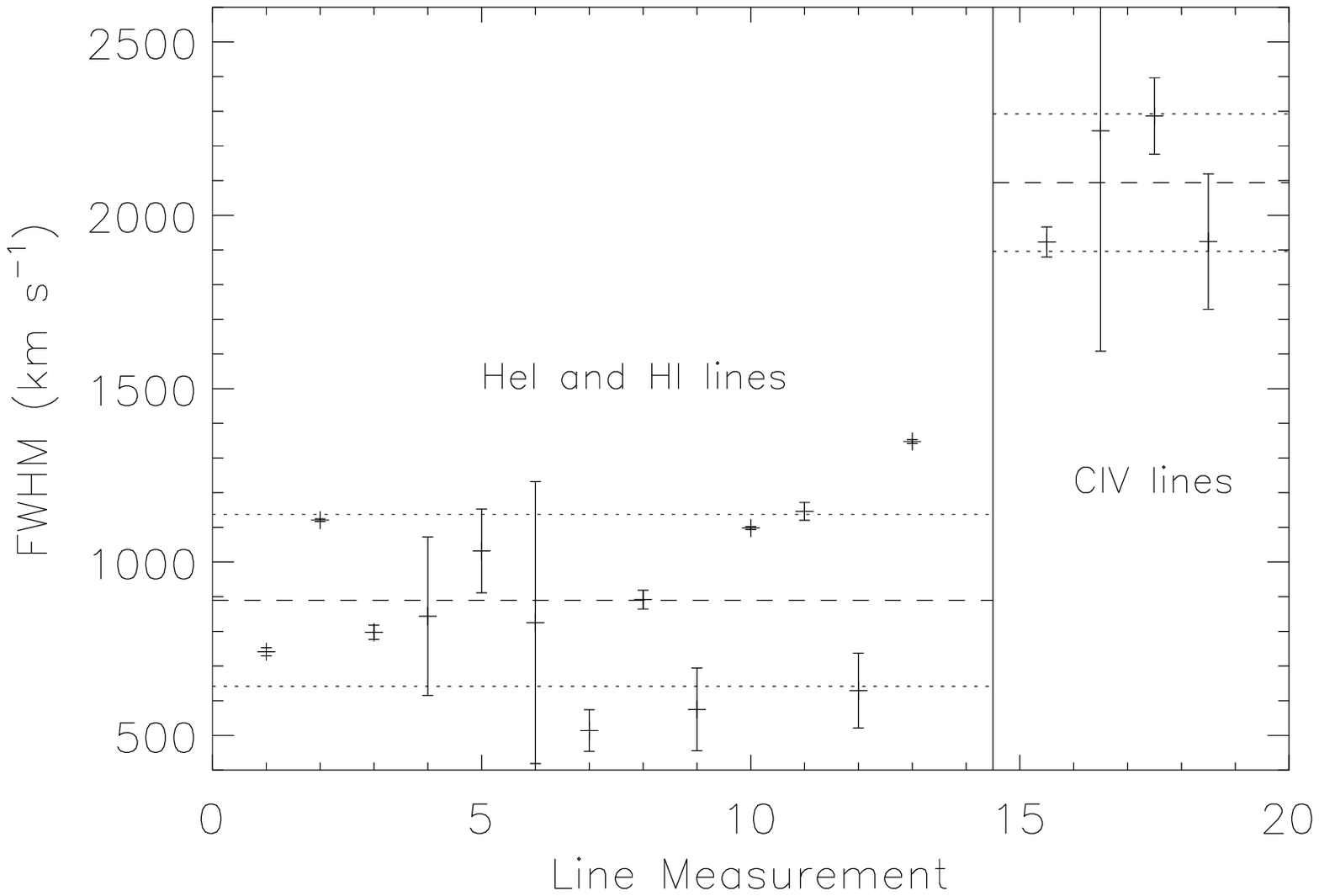}
\end{center}
\caption{The \WR line parameters.
{\it Left panel:}
The ratio of equivalent widths of various emission lines in
the optical and NIR spectra of \WRE: EW(SALT)/EW(AAT) for
$\lambda < 8000$~\AA; EW(ESO)/EW(AAT) for $\lambda > 9000$~\AA.
{\it Right panel:}
The full width at the half maximum (FWHM) of He\,{\sc i} 5876, 
6678, 10830~\AA~ and H$_{\alpha}$ (line measurement $< 14$); 
C\,{\sc iv} 5808~\AA~ and 7737~\AA~ (line measurement $> 15$).
The term `line measurement' (X-axis) has no specific meaning and is
used only for counting the derived FWHM values. 
}
\label{fig:ew}
\end{figure*}

The corresponding results for \WR are given in
Table~1. %Table~\ref{tab:wr48a_lines}. 
Since the equivalent widths do not
depend on the spectral resolution, their values can give an indication
for spectral changes. The EW ratios from the \WR spectra over a time
difference of about 20 years are shown in Fig.~\ref{fig:ew}. We see
that there is some indication that spectral lines (especially,
hydrogen-helium lines) in the optical are stronger in 2012
(the SALT observation), while the spectral lines in the NIR  were
stronger in 1993 (the AAT observation). The latter is likely due to 
appearance of some strong continuum in the NIR, probably a dust 
emission  in 2012.

It is interesting to note that the FWHM values of the optical-NIR
emission lines in \WR give some indication for a possible presence
of two gas flows different in kinematic sense. Namely, we see that all
the `cool' lines (He\,{\sc i} and H\,{\sc i} lines) are considerably 
narrower than those of high-excitation ionic species 
(e.g. C\,{\sc iv}). Figure~\ref{fig:ew}
(right panel) is an illustration of this difference. We note that 
the C\,{\sc iv} lines in \WR are as broad as in the WC8 star \WC
(Table~\ref{tab:wc8_lines}).
Then, could
it be that these `hot' lines belong to the WC8 star in \WR while the
narrower  helium-hydrogen lines (the `cool' lines) belong to its
companion star? We will return to this issue in
Section~\ref{subsubsec:wr48a_fit}.

\begin{table}
\setcounter{table}{1}
\caption{\WCE: emission lines}
\label{tab:wc8_lines}
\begin{center}
\begin{tabular}{rrr}
\hline
\multicolumn{1}{c}{} & \multicolumn{2}{c}{SALT}  \\
\multicolumn{1}{c}{$\lambda$} & 
\multicolumn{1}{c}{EW} & \multicolumn{1}{c}{FWHM}  \\ 
\hline
C\,{\sc iv} 4650  & 984.10$\pm$7.00  &   \\
C\,{\sc iii} 5696  & 455.30$\pm$2.50  &   \\
C\,{\sc iv} 5808  & 454.30$\pm$32.3  & 49.24$\pm$4.30  \\
      &                   & (2542$\pm$222)  \\
He\,{\sc i} 5876  & 110.90$\pm$7.00  &   \\
H$_{\alpha}$ 6563  & 144.20$\pm$8.40  &   \\
 C\,{\sc iii}$+$C\,{\sc iv} (?) $\sim$6597  &   &    \\
He\,{\sc i} 6678  & 302.60$\pm$0.68  &   \\
He\,{\sc i}, C\,{\sc iii} 7050  &   &   \\
C\,{\sc iii}, C\,{\sc iv} 7230  & 108.90$\pm$0.47  &   \\
C\,{\sc iv} 7737  &  94.84$\pm$0.34  & 59.24$\pm$0.23   \\
      &                  & (2296$\pm$9)  \\
\hline

\end{tabular}

\end{center}

Note -- Given are the line wavelength ($\lambda$) in \AA, the line
equivalent width (EW) in \AA, the line full width at the half maximum
(FWHM) in \AA, the FWHM values in units of \kms are given in 
parentheses.

\end{table}

\subsection{Interstellar absorption features}
\label{subsec:dibs}
For \WR and \WCE, we measured the equivalent widths of all the
identified interstellar absorption lines and diffuse interstellar 
bands in the SALT data. We just recall that the spectra of 
these two objects were obtained simultaneously. The results are 
given in Table~\ref{tab:dibs}. We see that the equivalent widths of
the interstellar absorption features are practically the same for both
objects. This indicates that \WR and \WC are located at the same
distance from us. Keeping in mind that \WC is a member of Danks 2, one
of the two central clusters in the G305 star-forming complex
\citep{davies_12}, we obtain an important confirmation 
that \WR is located at the distance of $\sim4$ kpc to Danks 1 and 2 
(e.g., \citealt{danks_83}; \citealt{davies_12}).
This conclusion is supported by the fact that `identical' EWs
values are measured for a near-by star S1 (D2-7) - another member of 
the open cluster Danks 2 (see Appendix~\ref{append:stars}).

\begin{table}
\caption{Interstellar absorption features}
\label{tab:dibs}
\begin{center}
\begin{tabular}{rrr}
\hline
\multicolumn{1}{c}{$\lambda$} & 
\multicolumn{1}{c}{\WR} & \multicolumn{1}{c}{\WC}  \\ 
\hline
Na\,{\sc I} 5889  & 1.26$\pm$0.08  & 1.37$\pm$0.08  \\
Na\,{\sc I} 5995  & 1.22$\pm$0.09  & 1.08$\pm$0.07  \\
6283  & 2.70$\pm$0.10  & 2.82$\pm$0.13  \\
6614  & 0.38$\pm$0.03  & 0.57$\pm$0.03  \\
K\,{\sc I} 7699  & 0.50$\pm$0.03  & 0.42$\pm$0.03  \\
8620  & 0.55$\pm$0.03  & 0.56$\pm$0.04  \\
\hline

\end{tabular}

\end{center}

Note -- Given are the absorption line identification or the DIB  
wavelength ($\lambda$) in \AA~ and its
equivalent width (EW) in \AA~ for \WR and \WCE, respectively.
\end{table}

\subsection{Global spectral fits}
\label{subsec:fits}
All the spectral fits were done by making use of the
Levenberg-Marquardt method for non-linear fitting (e.g., Section~15.5 in
\citealt{press_92}). As theoretical
spectra, we used the observed flux-calibrated spectra of stars of 
different spectral types taken from the spectral libraries (on-line 
data in SIMBAD) or from the archives of various observatories. 
In the spectral fits, the interstellar (ISM) extinction
was also taken into account by making use of the standard ISM
extinction curve of \citet{fitz_99}.

\subsubsection{Spectral fits to \WC}
\label{subsubsec:wc8_fit}
Since \WC (2MASS J13125770-6240599; 
WR48-2\footnote{Galachtic Wolf Rayet Catalogue;
\url{http://pacrowther.staff.shef.ac.uk/WRcat/index.php}}) 
was classified Wolf-Rayet star 
of the WC8 spectral type (\citealt{mauer_09}; \citealt{davies_12}),
we fitted its SALT spectrum with spectra of classical WC8 stars and
taking into account the interstellar extinction. Namely, we used the 
spectra of WR 135 (3140 - 7221 \AA) and WR 60 (3384 - 6867 \AA) from 
the atlas of WC-star spectra by \citet{torr_mass_87}. Since these
spectra are not corrected for the interstellar extinction, we first
dereddened them with $E_{b-v} = 0.34$ mag and 
$E_{b-v} = 1.40$ mag for the
WR 135 and WR 60, respectively 
($E_\mathrm{B-V} = 1.21E_{b-v}$,
see Section~8.2 and Table 28 in \citealt{vdh_01}).
We rebinned the
observed spectrum to increase its signal-to-noise and we rebinned the 
theoretical spectra correspondingly as well. We explored a rebinning of 
2, 5 and 10 \AA~ to check how the fit results depend on this parameter.
We recall that the SALT spectra were flux-calibrated but in relative
units (see Section~\ref{subsec:optical}) thus the free parameters in the
fits were the scaling factor for the theoretical spectrum at some 
fiducial wavelength and the optical extinction to \WCE. We note that
in the case under consideration the spectral scaling factor has no 
direct physical meaning. On the other hand, such fits allow us to
derive an estimate of the optical extinction and, as expected and also
proved in the fits, its derived value does not depend on what 
wavelength the scaling parameter was defined at.

\begin{figure*}
\begin{center}
\includegraphics[width=3.in, height=2.15in]{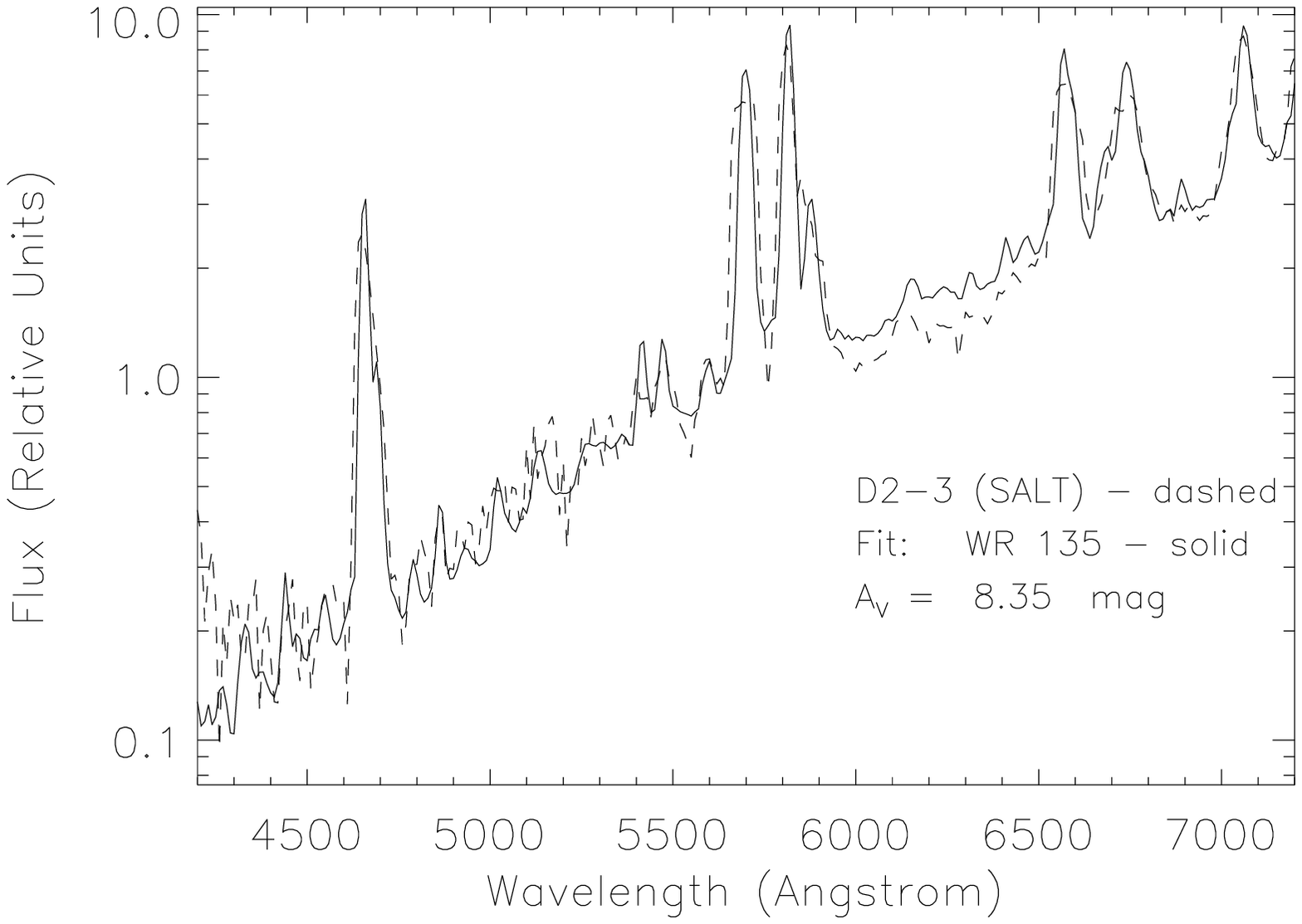}
\includegraphics[width=3.in, height=2.15in]{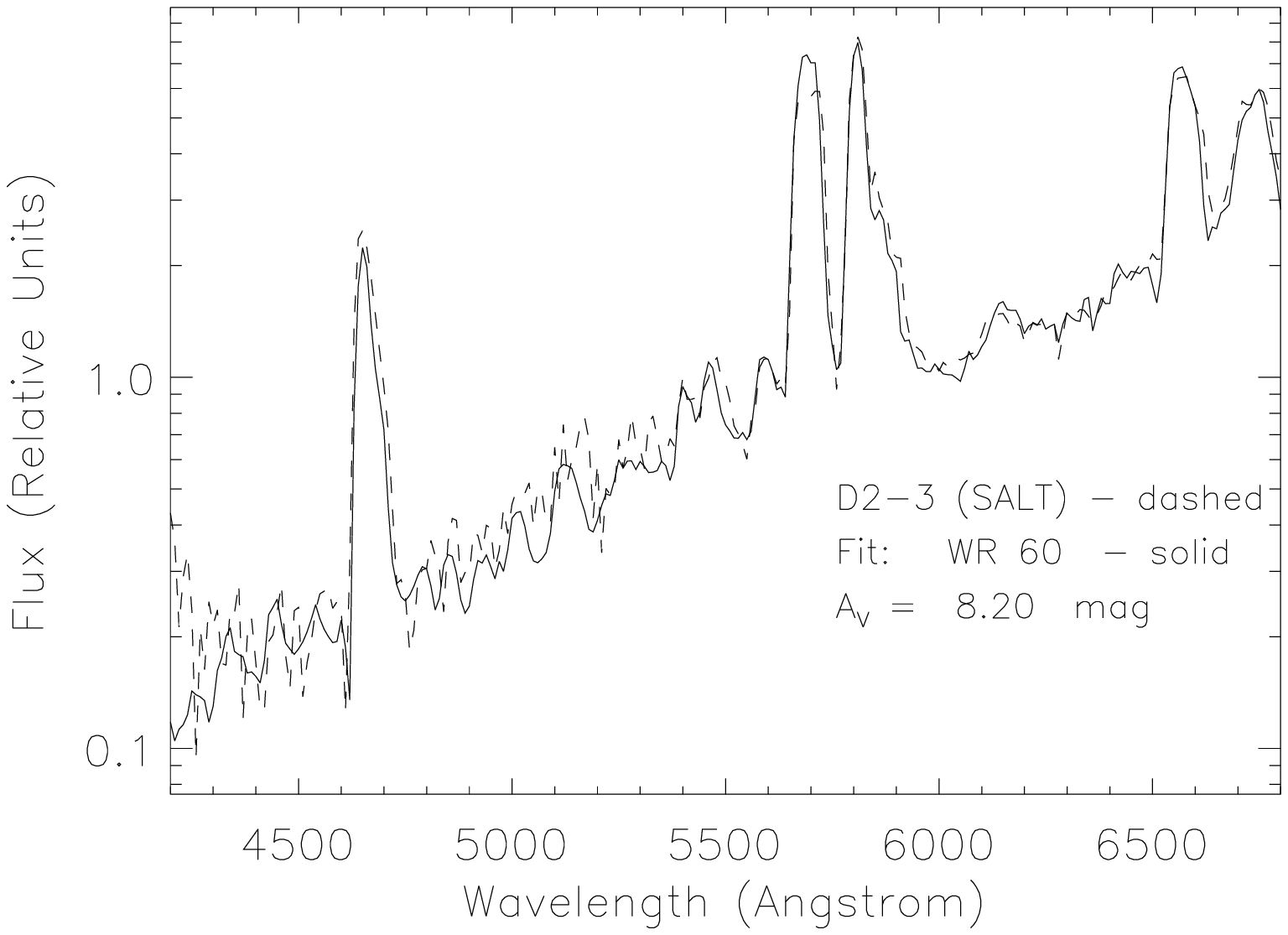}

\end{center}
\caption{
Fits to the SALT spectrum of \WCE. The `theoretical' spectra are 
those of WR 135 and WR 60 (both are  Wolf-Rayet stars of the WC8 
type). The bin size is 10~\AA.
}
\label{fig:wc8}
\end{figure*}

Figure~\ref{fig:wc8} presents some results from our fits. We see that 
the theoretical WC8 spectra nicely match the SALT spectrum of \WC which 
confirms its spectral classification as a WC8 Wolf-Rayet star. The 
derived optical extinction to this object is 
$E_\mathrm{B-V} = 2.67\pm0.03$ mag 
(the uncertainty corresponds to the variance of the mean of
the values derived in the fits for different spectral rebinning) 
or A$_{\mathrm{V}} = 8.28\pm0.09$ mag 
(A$_{\mathrm{V}} = 3.1E_\mathrm{B-V}$).
We note that the extinction to \WC is consistent with the value
we derived for another member of the open cluster Danks 2,
S1 (D2-7): $E_\mathrm{B-V} = 2.73$ mag 
(see Appendix~\ref{append:stars}).

\subsubsection{Spectral fits to \WR}
\label{subsubsec:wr48a_fit}
\WR was classified as Wolf-Rayet star of the WC8 type \citep{vdh_01}.
It has been suggested that \WR is a binary star and the relatively 
weak spectral lines in its spectrum hint on a presence of an additional
source of continuum emission that is attributed to the companion star
in the system. \citet{williams_12} gave 
arguments for the companion to be an emission-line star, O5Ve.
To explore this possibility in some detail, we carried out 
two-component spectral fits to the SALT and AAT spectra of \WR that
take into account the interstellar extinction. We note that for these
fits we constructed one common AAT spectrum. Namely, we adopted the
AAT blue-channel spectrum for wavelengths shorter than 5000~\AA~ and 
the AAT red-channel spectrum otherwise.
The details of the fitting procedure are as follows.

For the first component of the spectral fit, we chose the SALT 
spectrum of the \WC which allowed us to explore a larger spectral 
range (the SALT spectrum of \WC spans in wavelength almost to 9000~\AA 
~while that of WR 135 from the Torres \& Massey atlas spans only to 
7200~\AA). This spectral component was dereddened by 
$E_\mathrm{B-V} = 2.67$ mag to derive 
the `theoretical' spectrum of the first component in the fit. For the 
second component in the \WR fit, we used spectra from STELIB library 
\citep{leborgne_03} or from the archives of various observatories. 
We recall that all the `theoretical' spectra were flux-calibrated.  
The free parameters in the fit were the scaling factor for each 
spectral component and the interstellar extinction (both components 
were subject to common interstellar extinction). Thus the spectral 
fits of the SALT and AAT spectra of \WR provided us with an estimate 
of the interstellar extinction to this object and of the relative 
brightness of the stellar components. We represented the latter by 
the difference of the intrinsic V-magnitude of the WC8 star 
($V_{WC8}$) and its companion star ($V_C$): 
$\Delta V = V_{C} - V_{WC8}$.

\begin{figure*}
\begin{center}
\includegraphics[width=3.in, height=2.15in]{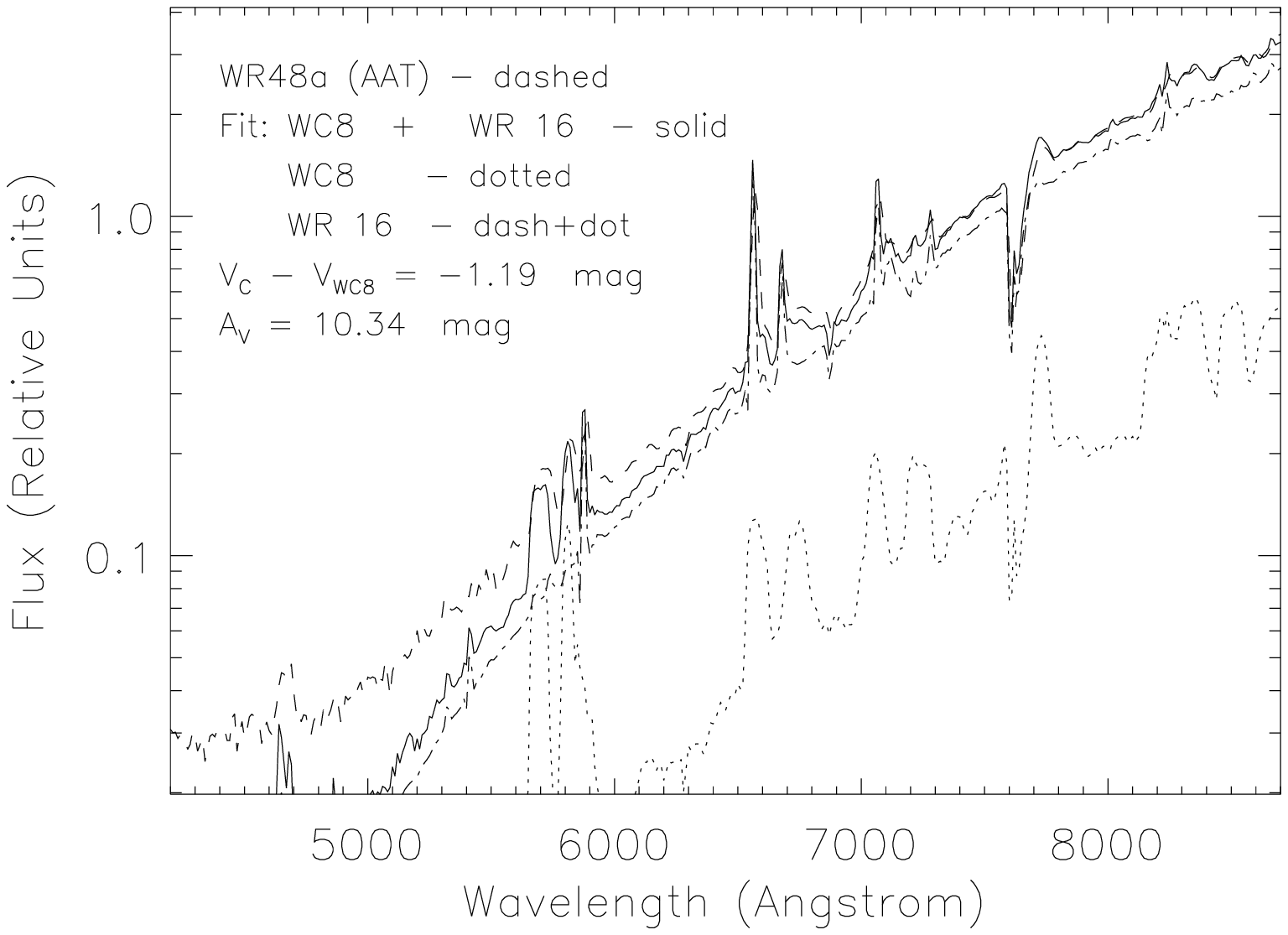}
\includegraphics[width=3.in, height=2.15in]{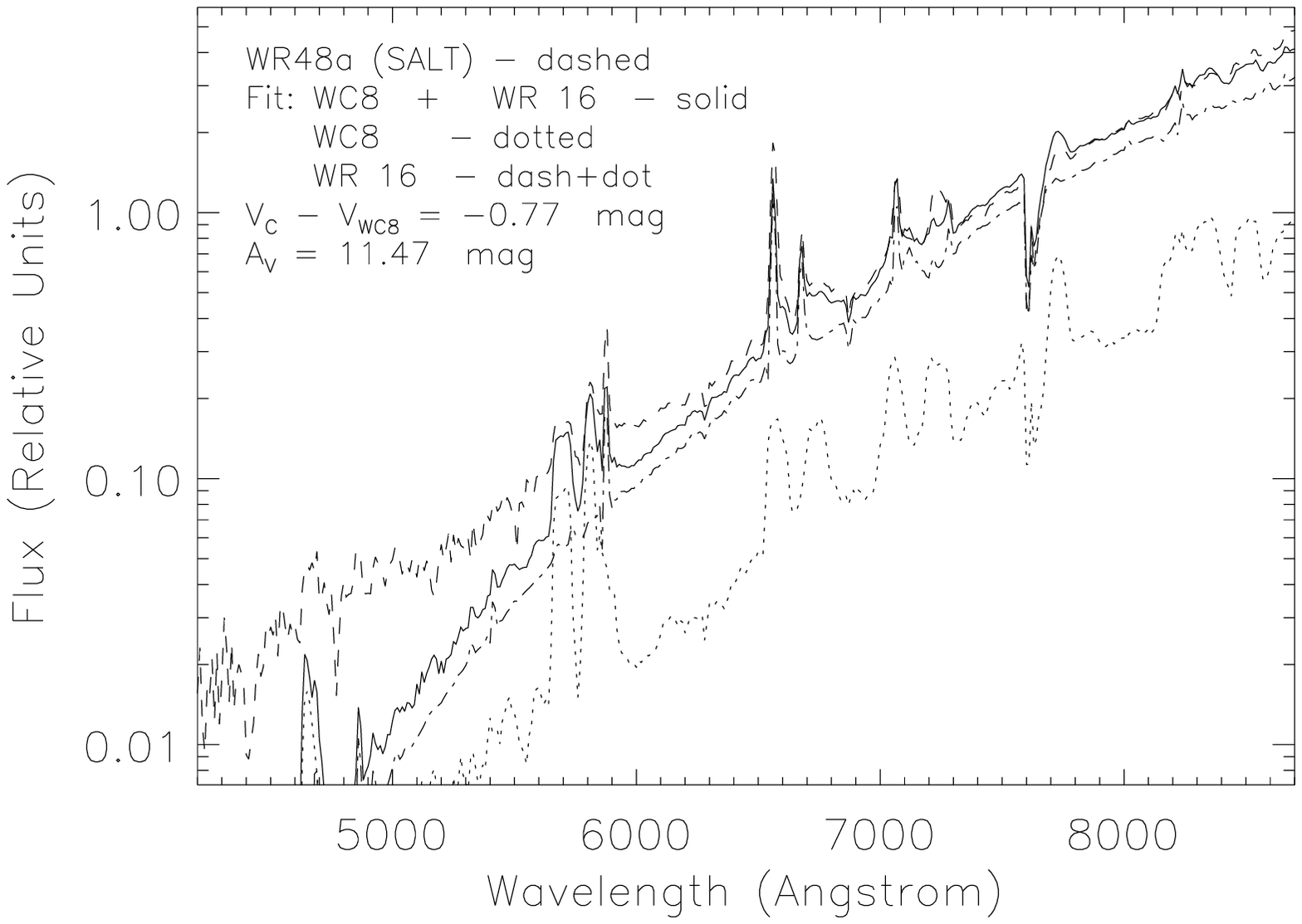}
\end{center}
\caption{
Two-component fits to the AAT and SALT spectra of \WRE. The first
component is the SALT spectrum of the \WC star and the second
component is the spectrum of WR 16 (a Wolf-Rayet star of the WN8h 
type). The bin size is 10~\AA.
}
\label{fig:wr16}
\end{figure*}

Since no spectral information is available about the presumable
stellar companion of the WC8 star in \WRE, we explored a range of
spectral types for it. For this, we made use of spectra from the 
STELIB library \citep{leborgne_03} and data from the European 
Southern Observatory (ESO) archives. We ran global two-component fits
to the SALT and AAT spectra of \WR with spectra of WR, LBV (luminous
blue variable), O, B and A
stars for the second component. We note that in these fits matching
the observed WC8 spectral features in the \WR spectra was not a
problem: these spectral lines are matched well if they are scaled
appropriately by the continuum of the second stellar component. The
basic challenge in these fits was to match the spectral lines of
helium and hydrogen (He\,{\sc i} 5876, 6678~\AA, H$_{\alpha}$~6563~\AA).
Our best fit with this respect was that with a spectrum of a WR star
of late WN type (WR 16; van der Hucht 2001) for the second spectral 
component. Figure~\ref{fig:wr16} presents the results from this 
two-component (WC8$+$WN8h) fit. The spectral fits with LBV, O, B or 
A-star spectrum for the second component did not give a good match to 
the helium and hydrogen lines (see Appendix~\ref{append:fits}). A few 
things are worth noting.

In overall, the interstellar extinction towards \WR in the V-filter 
is higher by 2-3 magnitudes than that towards the near-by WC8 star
\WCE. 
In all the fits, the \WR spectra at shorter wavelengths ($\lambda <
5500$~\AA) were not matched well by the two-component fit. We tried to
 improve the quality of the fit by adding another component (e.g., a
black-body or a power-law emission) but we could not succeed. We note
that the quality of the SALT and AAT spectra in that spectral range
is not very good and this might be the reason for such a discrepancy.
We thus think that new spectra with much better sensitivity are
needed to resolve this issue.
In the AAT spectrum, the companion star is about 3 times brighter than
the WC8 component ($\Delta V = V_{C} - V_{WC8} = -1.19$~mag). This is
consistent with the estimate by \citet{williams_12} based on the
ratio of the 5696-\AA~ and 5813-\AA~ lines in the AAT spectra of \WR
and of the WC8 standard WR 135.
Some spectral changes have taken place in the optical over a period of
time of about 19 years, that is between 1993 June (AAT) and 2012 May 
(SALT):
(a) the relative brightness of the companion decreased;
(b) the interstellar (or circumstellar) extinction towards \WR
increased.
We will further discuss these results in Section~\ref{sec:discussion}.

\subsection{Radio properties}
\label{subsec:radio_results}
As discussed in Section~\ref{subsec:radio}, we were able to derive the 
radio fluxes of \WR from the observations analysed inthis study.
The results are given in Table~\ref{tab:radio}. 
Although of limited spectral coverage (observed fluxes only at two
frequencies), these results indicate that the radio source in \WR is
likely of thermal origin: F$_{\nu} \propto \nu^{\alpha}, \alpha > 0$.
In fact, the derived power-law index (see Table~\ref{tab:radio}) 
is within 
$(1-2)\sigma$ from the canonical spectral index of $\alpha = 0.6$ for 
the thermal radio emission from the stellar wind in massive stars 
(\citealt{pf_75}; \citealt{wri_bar_75}). We recall that in their 
analysis of the radio data on the G305 star-forming region, 
\citet{hindson_12} also reported a detection of a thermal radio source 
($\alpha = 0.6$) associated with \WRE.
Comparing the radio fluxes from all these observations, two things
seem conclusive: (a) some variability in the radio emission of \WR 
is present; (b) the radio emission of \WR is  very likely of 
thermal origin.

\begin{table}
\caption{\WRE: radio fluxes}
\label{tab:radio}
\begin{center}
\begin{tabular}{lrrr}
\hline
\multicolumn{1}{c}{Date} & 
\multicolumn{1}{c}{4.8 GHz} & \multicolumn{1}{c}{8.64 GHz} &
\multicolumn{1}{c}{$\alpha$}\\
\hline
2000 Feb 27  & $3.94\pm0.21$  & $5.39\pm0.46$  & $0.53\pm0.17$ \\
2006 Dec 8   & $1.90\pm0.27$  & $3.64\pm0.18$  & $1.10\pm0.25$ \\
\hline

\end{tabular}

\end{center}

Note -- Given are the date of radio observation, the radio fluxes
with the $1\sigma$ error in mJy and the spectral index 
(F$_{\nu} \propto \nu^{\alpha}$).

\end{table}

For the case of thermal radio emission, we can estimate the ionized
mass-loss rate for an assumed constant velocity  smooth (no
clumps) wind using the result of \citet{wri_bar_75}: 
$\dot{M} = C_0 v_{\infty} S_{\nu}^{0.75} d_{kpc}^{1.5}$ \dotM, where
$C_0 = 0.095 \mu / [Z(\gamma g \nu)^{1/2}]$. Here $v_{\infty}$ is the
terminal wind velocity in \kms, $S_{\nu}$ is the radio flux in Jansky
at frequency $\nu$ in Hz, $d_{kpc}$ is the stellar distance in kpc,
$\mu$ is the mean atomic weight per nucleon, $Z$ is the rms ionic
charge, $\gamma$ is the mean number of free electrons per nucleon, and
$g$ is the free-free Gaunt factor (see equation 8 in \citealt{ab_86}
for a suitable approximation of the Gaunt factor). Since \WR is
a WC8 Wolf-Rayet star, we adopt: $Z = 1.4$, $\gamma = 1.3$, $\mu =
7.7$ typical for such an object \citep{vdh_86}, 
$v_{\infty} = 1700$\kms (the mean wind velocity for WC8 stars, see 
Table 27 in \citealt{vdh_01}), and $d_{kpc} = 4$ 
(see Section~\ref{subsec:dibs}). So, the derived radio fluxes correspond 
to a mass-loss rate of $\dot{M} = (3.7-6.4)\times10^{-4}$\dotM.
We only note that the \WR radio fluxes of 2.4 mJy (5.5 GHz) and 
2.9 mJy (8.8 GHz) from \citet{hindson_12} correspond to a mass-loss 
rate of $\dot{M} = (3.9-4.1)\times10^{-4}$\dotM.
We thus see that the level of thermal radio emission from \WR suggests
that this WC8 star has a very massive stellar wind.

\begin{figure}
\begin{center}
\includegraphics[width=\columnwidth]{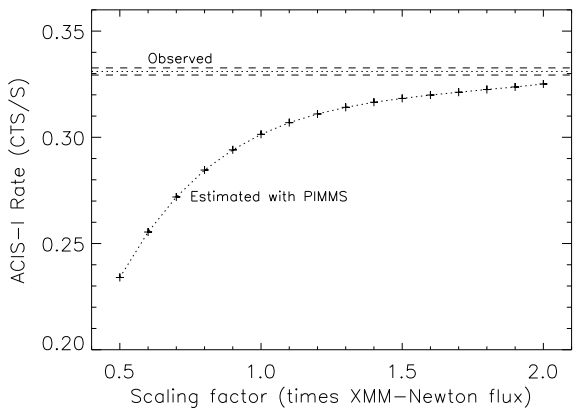}
\end{center}
\caption{The \ChandraE-ACIS-I count rates from the {\sc pimms}
simulations of the X-ray emission of \WRE. The count rates are
corrected for the pile up. The two dashed lines bracket the $1\sigma$
confidence interval of the observed count rate of \WR (\Chandra
ObsID 8922).
}
\label{fig:pimms}
\end{figure}

\subsection{X-ray light curve}
\label{subsec:xray_lc}
As reported by \citet{zhgsk_14}, \WR is a variable
X-ray source on a time scale of a few years. The \Swift observations,
taken on a quasi-monthly basis over a more than two-year 
period of time, provide additional information in this respect. We
extracted source and background spectra for 32 \Swift observations.
Unfortunately, the short exposure time (see Section~\ref{subsec:xrays}) 
and the decreased X-ray emission from \WR do not allow for carrying 
out a spectral analysis in detail. On the other hand, we used the data 
to construct an X-ray light curve. For comparison, we converted the 
\XMM and \Chandra fluxes into the \Swift reference system. This was 
done in the following manner.

For each of the \XMM and \ChandraE-HETG observations, we simulated a 
\Swift spectrum in \xspec by making use of the \Swift response 
matrix and the ancillary response files adopted in our study (see 
Section~\ref{subsec:xrays}). This provided the corresponding count rate in
the \Swift reference system. We examined all the ancillary response files
of the \Swift observations and found that the telescope effective area
has not changed considerably (e.g., by no more than a few percent) over 
the two-year monitoring of \WRE. We thus used the ancillary response
file for the \Swift observation (ObsID 00031900028), taken only two
days before the \ChandraE-HETG observation, in all the simulations of
this kind.
Also, the best-fit two-shock models that correspondingly matched the 
\XMM and \ChandraE-HETG spectra (see \citealt{zhgsk_11}, 2014) were 
used in the \xspec simulations.

As to the \ChandraE-ACIS-I observation, we recall that it is heavily
piled up (see Section~\ref{subsec:xrays}) and we made use of the 
\Chandra count rate prediction tool
{\sc pimms}\footnote{Portable, Interactive Multi-Mission Simulator; 
\url{http://asc.harvard.edu/toolkit/pimms.jsp}} to get an estimate of 
the flux level of \WRE. Namely, we adopted an absorbed two-temperature
thin-plasma model (model $apec$ in \xspecE) and fitted the X-ray
spectrum of \WR from the \XMM observation in 2008 Jan. We then ran a 
series of {\sc pimms} simulations adopting the \ChandraE-ACIS-I 
parameters for the AO8 observation period when the data were actually 
obtained. In these simulations, the observed flux was scaled while the 
shape of the spectrum was the same (e.i., no change of the plasma 
temperature and the X-ray absorption). We note that {\sc pimms} 
estimates the amount of pile-up in the data and thus provides a 
pileup-corrected count rate representative of its observed value. 
Figure~\ref{fig:pimms} presents the results from our {\sc pimms} 
simulations. A comparison with the observed count rate in 
\ChandraE-ACIS-I data illustrates that the X-ray flux of \WR in 
2008 Dec was at least as high as it was in 2008 Jan (the \XMM 
observation). For the X-ray light curve, we then assumed a value of 
the \ChandraE-ACIS-I flux equal to that in the \XMM observation.

\begin{figure}
\begin{center}
\includegraphics[width=\columnwidth]{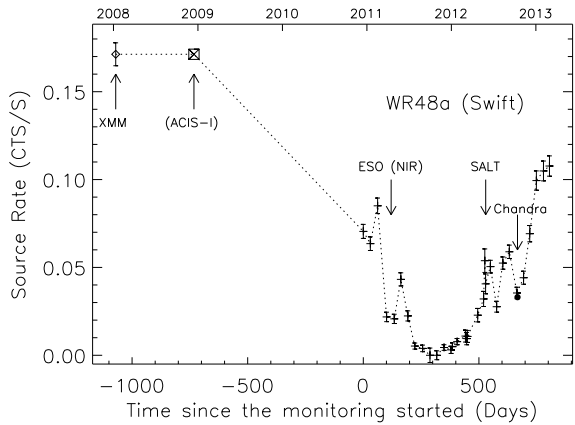}
\end{center}
\caption{The \Swift X-ray (0.5 - 10 keV) light curve of \WRE. Given on 
the axes are: the time since the \Swift monitoring started in 2010 Dec  
(X-axis; the calendar year of observations is shown on the upper 
X-axis) and the background-subtracted count rate (Y-axis). 
For comparison, the data points for the \XMM (2008 Jan), \Chandra
ACIS-I (2008 Dec) and \Chandra HETG (2012 Oct)  observations converted
into \Swift count rates (see text) are labelled `XMM' (diamond), 
`(ACIS-I)' (crossed square) and `Chandra' (filled circle), respectively. 
The ESO (NIR) and the SALT observation dates are marked as well.
}
\label{fig:swiftlc}
\end{figure}

\begin{figure}
\begin{center}
\includegraphics[angle=-90., width=\columnwidth]{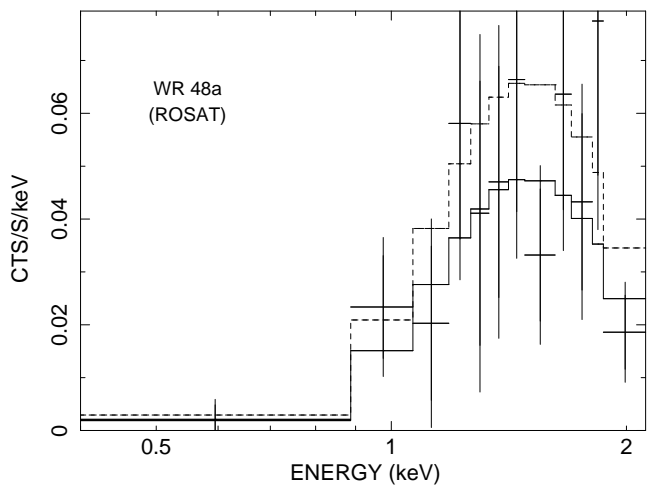}
\end{center}
\caption{The background-subtracted \Rosat spectrum of \WR rebinned to
have a minimum of 10 counts per bin. Overlaid are: 
(a) the two-shock model that perfectly fits the 2008-Jan \XMM spectra 
of \WR \citep{zhgsk_11} - dashed line;
(b) the same model but with varied normalization parameter to
match the flux level of the \Rosat spectrum (see text) - solid line.
}
\label{fig:rosatspec}
\end{figure}

Figure~\ref{fig:swiftlc} presents the background-subtracted light 
curve of \WR `expanded' with the \XMM and \Chandra data points. We 
note a very good correspondence between the fluxes (count rates) 
from the \Chandra-HETG observation on 2012 Oct 12 and the \Swift 
observation on 2012 Oct 10. This in turn means that the point-to-point
fluctuations in the \Swift light curve of \WR might be real and not an
instrumental artefact.

Finally, it is interesting to check the level of X-ray emission from
\WR on even a longer time scale by considering the \Rosat observations
carried out in 1997 Feb. To do so, we used the two-shock model that
perfectly matched the X-ray emission from \WR as of 2008 Jan (the \XMM
observation; \citealt{zhgsk_11}).
Figure~\ref{fig:rosatspec} shows the total \Rosat spectrum overlaid
with the two-shock model for the \XMM observation. We see that the
\Rosat data are consistent with this model which also means that the
X-ray emission of \WR in 1997 Feb was about the same as of 2008 Jan 
(the \XMM observation). In fact, if we kept the spectral shape
unchanged (e.i., plasma temperature and X-ray absorption fixed) and
varied only the normalization parameter of the model, the best-fit
\Rosat flux was about 70\% of that in the \XMM data but the difference
is within the $1\sigma$ confidence interval for the \Rosat flux. 

Thus from the X-ray light curve of \WRE, we have 
indications of a high and stable X-ray flux for a period of 10 years 
or so (from the \RosatE, 1997 Feb, through the \XMME, 2008 Jan , to 
the \ChandraE-ACIS-I, 2008 Dec , observation). Also, we see that
recently the X-ray emission from \WR decreased to a much lower level 
over a period of one-to-two years (2010 Dec - 2013 Feb). In
Section~\ref{sec:discussion}, we will discuss some implications from such 
a long-term X-ray variability in \WRE.

\section{Discussion}
\label{sec:discussion}

The basic results from our analysis of the new (optical and NIR) and
archive (optical, radio, X-rays) data on the dusty WR star \WR are as
follows.

{\it
In the optical,} the spectrum of \WR is acceptably well represented by
a sum of two spectra: of a WR star of the WC8 type and of a WR star of 
the WN8h type. An indication for a presence of two different gas flows
(stellar winds) is also found from the quite different widths of the
emission lines of `cool' species (He\,{\sc i} and H\,{\sc i}) and of 
`hot' species (C\,{\sc iv}): $\sim 900$~vs. $\sim 2100$\kms, 
respectively.
Some variability of the optical emission of \WR is established from
the comparison of the optical spectra of \WR obtained in 2011-2012 and 
in 1993.  This is indicated by the change of the equivalent widths of 
some emission lines, the relative brightness of the WC8 and WN8h 
spectral components and the optical extinction to this object.

{\it
In the radio,} \WR is  very likely a thermal source and also 
shows some variability on a timescale of a few years or so. The 
level of the radio emission suggests a relatively high mass-loss rate 
of this dusty WR star 
($\dot{M} \approx \mbox{a few} \times 10^{-4}$\dotM).

{\it
In the X-rays,} \WR is known to be a variable source on a timescale of
a few years \citep{zhgsk_14} and its variability on a shorter
timescale (within 1-3 years) is now established as well.

We recall that \WR shows recurrent dust formation on a 30-year
timescale which suggests that it is a long-period colliding wind
binary \citep{williams_12}. The importance of CSWs for the physics
of this object is also supported by its high X-ray luminosity
\citep{zhgsk_11} and by the fact that the forbidden line of Si XIII 
in its X-ray spectrum is strong and not suppressed \citep{zhgsk_14}. 
The latter indicates that a rarefied 10-30 MK plasma forms far from 
strong sources of ultra-violet emission, most likely in CSWs in a wide
binary system.
It is thus interesting to `project' our
results onto this physical picture and see what new they could add or
help us constrain in it.

The basic features of CSWs in episodic dust makers are best
illustrated by the prototype for these objects, the WR$+$O binary 
WR 140. 
The maximum of the NIR emission, and the sudden onset of dust 
formation, occurs at or immediately after the orbital phase 
corresponding to the periastron passage \citep{williams_90}.
The X-ray emission from WR 140 is strong and variable
(\citealt{williams_90}; \citealt{zhsk_00}; \citealt{po_05}).
The non-thermal radio emission of this system is strong and variable 
(\citealt{williams_90}; \citealt{whi_be_95}; \citealt{do_05}).
All these characteristics are result from CSWs in a wide binary system 
with highly elliptical orbit \citep{williams_90}.

Apart from the NIR variability, the X-ray properties of \WR play a key
role for the presumable CSW picture in this object. In fact, \WR is
the X-ray most luminous WR star in the Galaxy after the black hole
candidate Cyg X-3, {\it provided it is physically associated with the
open clusters Danks 1 and Danks 2 at a distance of $d \sim 4$~kpc}
\citep{zhgsk_11}. We now have a good argument that this is indeed
the case. Our analysis of the SALT spectra of \WR and the nearby WR
star \WCE, which is a member of Danks 2 (e.g. \citealt{davies_12}), 
showed that both objects are located at about the same distance from 
us (see Section~\ref{subsec:dibs}). Thus, the very high X-ray luminosity of
\WR presents a solid support for the CSW picture in this object.

However, we have to keep in mind that
there are some important observational findings that pose 
problems for the standard CSW paradigm in the case of \WRE.

As noted by \citet{williams_12}, the infrared light curve of \WR 
differs appreciably from that of canonical dust-makers as WR 140. 
Namely, the infrared emission of the latter surges to a maximum in a
short period of time (due to the sudden onset of dust formation)  and 
then decreases much more slowly to its minimum (see Fig.2 in
\citealt{williams_90}). Opposite to this, the infrared light curve of 
\WR gradually changes between its minimum and maximum values (see 
Fig.3 in \citealt{williams_12})
and even shows some `mini erruptions' as observed in another 
episodic dust maker WR 137 \citep{williams_01}.

On the other hand, the sudden onset of dust formation is associated
with the periastron passage in a wide bianry system with elliptical
orbit \citep{williams_90}. This means that the maximum of the 
infrared emission occurs near the orbital phase of the minimum binary 
separation. Interestingly, the \ChandraE-HETG X-ray observation of
\WR was carried out near the maximum of the infrared emission of this
dusty WR star. The analysis of these data 
{\it  along the lines of the adiabatic CSW picture (appropriate for
wide WR binaries)} 
showed that the binary separation at the moment of the
\Chandra observation (2012 Oct 12) was {\it larger} than that at the 
moment of the \XMM observation (2008 Jan 9) of \WR which was taken 
approximately 5 years earlier \citep{zhgsk_14}. Thus, the maximum
of the infrared emission from \WR does not seem to be related to the
minimum binary separation (the periastron passage) in this presumable
wide binary system. It may well be that the mechanism for infrared
emission in \WR is different from that in canonical dust-makers as WR
140.

Since \WR is very likely a thermal radio source (see
Section~\ref{subsec:radio_results}),
the lack of non-thermal radio (NTR) emission from this object 
is another atypical characteristic of a
possible wide CSW binary. We recall that in general the CSW binaries are 
NTR sources \citep{do_00}. We could think of at least two
possible reasons for such a discrepancy. 

First, the NTR source could
be heavily absorbed by the massive wind(s) in the binary if it were
located deep in the radio photosphere. We note that for the typical WC
abundances the mass-loss rate
of the stellar wind(s) in \WR is very high
$\dot{M} = (3.7-6.4)\times10^{-4}$\dotM
(see Section~\ref{subsec:radio_results}). It is very high even if we
assume the typical WN abundances \citep{vdh_86} or the
solar abundances \citep{an_89} for the stellar wind(s):
$\dot{M} = (3.0-5.2)\times10^{-4}$\dotM  (WN);
$\dot{M} = (1.0-1.7)\times10^{-4}$\dotM  (solar).
For these values of the mass-loss rate and using eq.(11) from
\citet{wri_bar_75}, we derive that the radius of the radio photosphere
(defined for a radial optical depth from infinity of 0.244) is in the
range 95 - 203 AU for the radio frequencies of the \WR observations.
In turn, the radius of the radial optical depth of unity is 23 - 50
AU. Thus, if the NTR source in \WR is located in the stellar wind
deeper than that radius, we may have missed its detection. 

Second, could it be that {\it no} NTR source is physically present in 
\WRE? In general, we need two ingredients for generating non-thermal 
radio emission: relativistic electrons and a magnetic field. The
relativistic electrons are assumed to be born in strong shocks through
the mechanism of diffusive shock acceleration. This mechanism also
requires a presence of a magnetic field (e.g., \citealt{bell_78}). 
Strong shocks are definitely present in a colliding wind binary but if 
there is no magnetic field in \WRE, then the generation of NTR emission 
would not be possible. If future, multi-frequency, radio observations
establish with certainty that \WR has no {\it intrinsic} NTR emission, 
this will be a very important finding. We note that the spectral fits 
to the optical emission of \WR indicated that its binary 
components are evolved stars, namely, a WC and a WN (or a LBV) star 
(see Section~\ref{subsubsec:wr48a_fit}). Thus, the lack of NTR emission 
will be a sign that no magnetic field is present in such an 
evolutionary phase of massive stars. This will also mean that in all 
the WR$+$O binaries that are source of NTR emission the O star 
provides the magnetic field that is needed for the relativistic 
electron production in strong shocks.

Thus, we see that although there are observational facts that strongly
indicate the presence of colliding stellar winds in \WRE, there is
also a number of observational facts which indicate that the physical
picture in this dusty WC star might be more complex than that.

Along these lines, we mention the relative variability between the WC
and the WN components as revealed by the change of the equivalent
widths of some strong optical lines and by the spectral fits to the
optical emission of \WR (see Section~\ref{subsec:lines} and
~\ref{subsubsec:wr48a_fit}). This spectral variability is also
accompanied by a change in the optical extinction to this object (see
Section~\ref{subsubsec:wr48a_fit}). 
We believe that this variable optical extinction is of local origin.
More specifically, we mean the additional extinction to \WR compared 
to that to \WC (the near-by member of the open cluster Danks 2): the
A$_{\mathrm{V}}$ value for the former is by 2-3 mag higher than for 
the latter. Interestingly, the higher local extinction is observed 
near the maximum of the infrared emission from \WR (see Fig.3 in
\citealt{williams_12}): the SALT spectrum (2012 May 27) shows higher 
optical extinction than the AAT spectra (1993 June 21). The local 
dust formation, with a variable rate of dust production, is a 
possible mechanism that could explain these results. Then, the origin 
of the dust in \WR is a key issue: does the circumstellar dust form 
in CSWs in a wide binary with elliptical orbit or in a different 
place?

We already discussed some observational findings that strongly support
the standard CSW picture in \WR and some that pose problems for its
validity. In this respect, we recall that the highest rate of dust 
formation in CSWs (maximum infrared emission) is associated with the 
periastron passage (minimum binary separation) in a wide binary with 
elliptical orbit. However, the analysis of the X-ray data of \WRE, 
obtained with \Chandra near the maximum of the infrared emission, 
showed that the binary separation was not near its minimum value 
\citep{zhgsk_14}. Thus, could it be that the dust formation is 
related to some LBV-like activity, as the latter might be indicated 
by the relative 
variability between the WC and the WN components in the optical 
spectrum of \WRE? Could \WR be similar to the binary (or triple) 
system HD 5980 in the Small Magellanic Cloud having a LBV object as one 
of its component and showing a 40-year variability cycle (see 
\citealt{gloria_10} and the references therein)? We only note 
that dust exists in the LBV environment (e.g., \citealt{mcgreg_88}) 
and the presence of a WC star (a carbon-rich object) in \WR may 
additionally facilitate the dust production in this system.

The high mass-loss values, deduced from the analysis of the radio
data, hint on something unusual in the dusty WR star \WRE. Similar
indication comes also from the X-ray light curve obtained with \Swift
and supplemented by data from \ChandraE, \Rosat and \XMM (see 
Section~\ref{subsec:xray_lc} and Fig.\ref{fig:swiftlc}). A plausible
explanation of the long-lasted (300-500 days; $\sim 1-1.5$~years) 
minimum of the X-ray emission from \WR could be that it is due to
an occultation of the CSW X-ray emission by a very massive stellar
wind (or a recently expelled LBV `shell'). 

As discussed by \citet{zhgsk_14}, 
the increased X-ray absorption near the maximum of the
infrared emission from \WR may indicate that the orbital inclination
of this binary system is close to 90 degrees. In such a case, we may
expect to detect a primary maximum of the X-ray absorption when the
WC8 component of the system is located between the CSW region and the
observer. We note that the available data show that \WR was bright in
X-rays on a considerably long timescale, e.g. 10-12 years or so 
(see Section~\ref{subsec:xray_lc}), but the X-ray observations were scarce
and we may have missed that primary maximum of the X-ray absorption.

On the other hand, if the increased X-ray absorption is related to
some LBV-like activity in \WRE, having a quasi-period of 30-32 years,
then the high X-ray attenuation will be observed {\it only} during the
time when a new LBV `shell' is expelled.

We thus believe that future observations in various spectral domains 
will be very helpful for our understanding of the physical picture in 
this fascinating object \WRE. Good quality, high resolution, optical 
spectra, taken regularly on approximately one-year basis, will allow 
us to constrain the physical properties (wind parameters, luminosity,
spectral type etc.) of the stellar components in this dusty Wolf-Rayet 
star. These observations should be supplemented by optical photometry 
as well. Multi-frequency radio observations with high sensitivity 
will help us reveal if a non-thermal radio source exists in \WR or the 
system is just a thermal radio source as the current observations show. 
Deep high resolution X-ray spectra, taken with \ChandraE, are needed 
to follow the changes of the physical properties of the hot plasma 
that likely originates from CSWs in \WRE. In addition, regular 
observations, as the ones carried out with \SwiftE, are important as 
well for obtaining crucial information on the X-ray variability of 
\WRE, correspondingly on the global geometry of of this presumable 
binary (or of a higher hierarchy) system.

The data from such comprehensive observational studies can be used
to model the global properties of the dusty Wolf-Rayet star \WR in
considerable detail. This modelling can be carried out by making use
of the stellar atmosphere models (for the optical emission), 
hydrodynamic models of colliding stellar winds (X-ray and non-thermal 
radio emission, if the latter is detected) and detailed physical 
models of dust emission (infrared 
emission). Such a global analysis of the physical characteristics of
\WR will allow us to thoroughly test our understanding of the physics
of massive stars.

\section{Conclusions}
\label{sec:conclusions}

In this paper, we presented an analysis of the \WR emission in 
different spectral domains. The basic results and conclusions from 
such a multi-wavelength view on this dusty Wolf-Rayet star are as 
follows.

(i) The optical spectrum of \WR is acceptably well represented by
a sum of two spectra: of a WR star of the WC8 type and of a WR star of
the WN8h type.
The comparison of the \WR spectra from 2011-2012 and 1993 reveals that
the optical emission is variable.
This is indicated by the change of the equivalent widths of some
emission lines, the relative brightness of the WC8 and WN8h spectral
components and the optical extinction to this object.

(ii) Analysis of the interstellar absorption features in the spectra
of \WR and the near-by stars \WC (2MASS J13125770-6240599) and S1 
(D2-7; 2MASS J13125864-6240552) showed that these objects are at about 
the same distance from us. Keeping in mind that \WC and S1 are members 
of the open cluster Danks 2, we conclude that \WR is located at the 
distance of $\sim4$ kpc to Danks 1 and 2 (e.g., \citealt{danks_83};
\citealt{davies_12}).

(iii) \WR is  very likely a thermal radio source and shows some 
variability on a timescale of a few years or so. The level of the 
radio emission suggests a relatively high mass-loss rate of this dusty 
WR star ($\dot{M} \approx \mbox{a few} \times 10^{-4}$\dotM).

(iv) Analysis of the data from different X-ray observatories
(\ChandraE, \RosatE, \XMME) revealed that \WR was a very bright X-ray 
source for an appreciably long period of time (e.g. 10-12 years or so). 
Observations with \Swift have now established its variability on a 
shorter timescale (within 1-3 years).

(v) The results from our multi-wavelength view on the dusty Wolf-Rayet
star \WR in conjunction of the ones from the in-detail studies of its
infrared emission \citep{williams_12} and of its X-ray spectra
(\citealt{zhgsk_11}, 2014) show that colliding stellar winds likely
play a very important role in the physics of this object. However,
the global physical picture in \WR may not be that simple and
some LBV-like activity could not be excluded as well.

(vi) Future observations of \WRE, taken on a regular time-basis along
the entire spectral domain: from radio to X-tay, will be very helpful 
for our understanding of the physics of this fascinating Wolf-Rayet
star. In-detail modelling of this global data set will allow us to 
thoroughly test our understanding of the physics of massive stars.

\section{Acknowledgments}
This paper uses observations made at the South African Astronomical
Observatory (SAAO). The corresponding data were obtained with the 
Southern African Large Telescope (SALT) and the Radcliffe telescopes.
It is also based on observations made with NTT telescope at the La 
Silla Observatory, ESO, under programme ID 087.D-0490A.
The paper is also partly based on data from the ESO Science 
Archive Facility (under request numbers 110999, 111361 and 111364)
and from the Anglo Australian Telescope data archive.
The Australia Telescope Compact Array is part of the Australia
Telescope which is funded by the Commonwealth of Australia for 
operation as a National Facility managed by CSIRO.
This research has made use of data and/or software provided by the
High Energy Astrophysics Science Archive Research Center (HEASARC),
which is a service of the Astrophysics Science Division at NASA/GSFC
and the High Energy Astrophysics Division of the Smithsonian
Astrophysical Observatory.
This research has made use of the NASA's Astrophysics Data System, and
the SIMBAD astronomical data base, operated by CDS at Strasbourg,
France.
J.B. and R.K. are supported by the Ministry of Economy, Development, 
and Tourism’s Millennium Science Initiative through grant IC12009,
awarded to The Millennium Institute of Astrophysics (MAS) and Fondecyt 
Reg. No. 1120601 and  No. 1130140.
The authors are grateful to the referee, Peredur Williams, for the 
valuable comments and suggestions.

{}

\appendix

\section{The SALT spectra of other stars}
\label{append:stars}

As was noted in Section~\ref{subsec:optical}, the long-slit
mode of the SALT RSS spectrograph permitted us to obtain spectra of
four additional stars S1, S2, S3 and S4, simultaneously with the
spectra of \WR and \WCE. 
The SALT spectra of CPD-62~3058, a B star in close vicinity to \WRE,
were obtained with the same setup on 2012 May 24. All
these objects are marked in Fig.~\ref{fig:salt_slit} and their
spectra were reduced and analysed in the same way as
described in Section~\ref{subsec:optical}. Some information
for these objects is given in
Table~\ref{obj_list}. The EWs of the interstellar lines and DIBs
in their spectra are presented in Table~\ref{tab:add_dibs}.

\begin{table}
\caption{List of other stars observed with SALT}
\label{obj_list}
\begin{center}
 \begin{tabular}{lllc}
\hline
\multicolumn{1}{c}{Object} & \multicolumn{1}{c}{2MASS} & 
\multicolumn{1}{c}{V} & \multicolumn{1}{c}{Spectral} \\
\multicolumn{1}{c}{name} & &  
\multicolumn{1}{c}{mag} & \multicolumn{1}{c}{type$^\mathrm{c}$} \\
\hline
B (CPD-62~3058) & J13123927$-$6243048 & 10.77$^\mathrm{a}$ & B3 \\
S1 (D2-7) & J13125864$-$6240552 & 15.91$^\mathrm{b}$ & O8 \\
S2 & J13124234$-$6242379 & 15.12$^\mathrm{b}$ & G0 \\
S3 & J13124037$-$6242518 & & \\
S4 & J13123545$-$6243221 & 
   & 
\\
\hline
\end{tabular}
\end{center}

Note --
 $^\mathrm{a}$ \citet{hog_00};
 $^\mathrm{b}$ \textsc{nomad};
 $^\mathrm{c}$ As derived here from the spectral fits.

\end{table}

\begin{figure}
\begin{center}
\includegraphics[width=\columnwidth,clip=true]{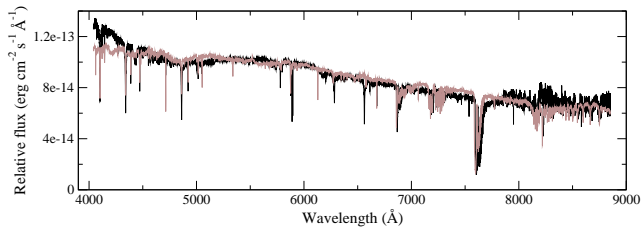}
\end{center}
\caption{The STELIB library \citep{leborgne_03} spectrum (in light
brown)
of HD 271163 (B3Ia), reddened by $E_\mathrm{B-V}=0.74$~mag and fitted
to the SALT spectrum (in black) of CPD-62 3058.}
\label{fig:cpd1}
\end{figure}

\begin{figure}
\begin{center}
\includegraphics[width=\columnwidth,clip=true]{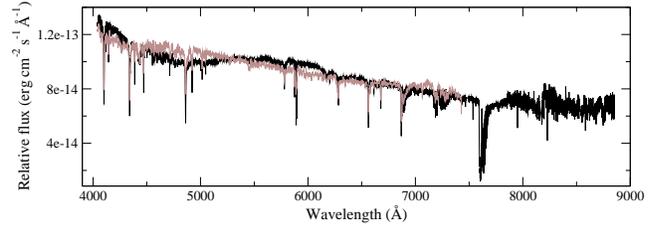}
\end{center}
\caption{The \citet{jacoby_84} library spectrum (in light brown) of 
HD 166125 (B3III), reddened by $E_\mathrm{B-V}=0.76$~mag and fitted to 
the SALT spectrum (in black) of CPD-62 3058.}
\label{fig:cpd2}
\end{figure}

\begin{figure}
\begin{center}
\includegraphics[width=\columnwidth,clip=true]{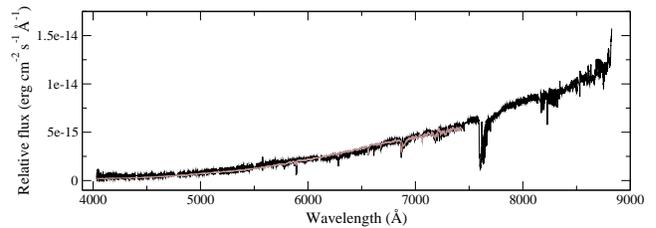}
\end{center}
\caption{The \citet{jacoby_84} library spectrum (in light brown) of 
HD 236894 (O8V), reddened by $E_\mathrm{B-V}=2.73$~mag and fitted to 
the SALT spectrum (in black) of S1.
}
\label{fig:cpd3}
\end{figure}

\begin{figure}
\begin{center}
\includegraphics[width=\columnwidth,clip=true]{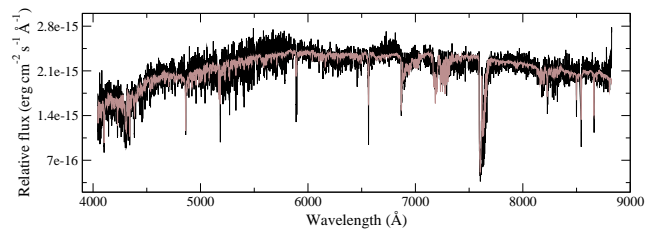}
\end{center}
\caption{The STELIB library \citep{leborgne_03} spectrum (in light
brown) of HD 63077 (G0V), reddened by $E_\mathrm{B-V}=0.46$~mag and 
fitted to the SALT spectrum (in black) of S2.
}
\label{fig:cpd4}
\end{figure}

An attempt to fit the observed spectra and to estimate the
interstellar extinction was made by the use of the tasks 
{\sc fitgrid} and {\sc fitspec} in the {\sc stsdas 
synphot}\footnote{{\sc stsdas} is a product of the
Space Telescope Science Institute, which is operated by AURA for
NASA. Also, 
see~
\url{http://www.stsci.edu/institute/software_hardware/stsdas/synphot/SynphotManual.pdf}}
package. First, we ran {\sc fitgrid} to find the best match to each 
of our spectra from the libraries of standard spectra. We then used 
{\sc fitspec} with two free variables, {\sl vegamag} in V passband 
and the $E_\mathrm{B-V}$ colour excess, for a more refined fit to our 
spectra. We note that the {\sl vegamag} parameter is just a scaling 
factor for the fit and has no specific physical meaning since the SALT 
spectra are flux-calibrated in relative units (see 
Section~\ref{subsec:optical}). The templates used were from the 
libraries of \citet{jacoby_84} and \citet{leborgne_03}.
The spectra of  \citet{jacoby_84} cover the wavelength
range 3510-7427\,\AA\ at a resolution of approximately 4.5\,\AA\  and
are included in the STSDAS package. The spectra in the STELIB 
\citep{leborgne_03} cover the range 3200-9200\,\AA\ with a resolution
$\lesssim3$\,\AA. The fits to the spectra of CPD-62~3058, S1 and S2 are
shown in Figs. ~\ref{fig:cpd1}, \ref{fig:cpd2}, \ref{fig:cpd3} and
\ref{fig:cpd4}. The spectra of the stars S3 and S4 were not fitted
because of the low S/N ratio.

\begin{table*}
\caption{Interstellar absorption features in the spectra of 
CPD-62~3058, S1, S2, S3 and S4.}
\label{tab:add_dibs}
\begin{center}
\begin{tabular}{rrrrrr}
\hline
\multicolumn{1}{c}{$\lambda$} & \multicolumn{1}{c}{CPD-62~3058} & 
\multicolumn{1}{c}{S1} & \multicolumn{1}{c}{S2} & 
\multicolumn{1}{c}{S3} & \multicolumn{1}{c}{S4} \\ 
\hline
5780  & $0.57\pm0.01$  &  &  &  &  \\
5797  & $0.16\pm0.01$  &  &  &  &  \\
NaI 5889  & $0.72\pm0.01$ & $1.13\pm0.03$ & $1.03\pm0.05^\mathrm{a}$ & 
            $2.14\pm0.09^\mathrm{a}$ & $1.34\pm0.09^\mathrm{a}$ \\
NaI 5995  & $0.62\pm0.01$  & $1.04\pm0.04$ & $0.77\pm0.04^\mathrm{a}$ & 
            $1.78\pm0.10^\mathrm{a}$ & $1.12\pm0.11^\mathrm{a}$ \\
6283  & $1.56\pm0.01$  & $2.83\pm0.06$ &  &  &  \\
6614  & $0.18\pm0.01$  & $0.44\pm0.03$ &  &  &  \\
KI 7699  & $0.20\pm0.03$  & $0.36\pm0.03$ & $0.36\pm0.07$ & $0.60\pm0.13$ &  \\
\hline

\end{tabular}
\end{center}

Note -- Given are the absorption line identification or the DIB  
wavelength ($\lambda$) in \AA~ and its
equivalent width (EW) in \AA~ for CPD-62~3058, S1, S2, S3 and S4, 
respectively.\\
$^\mathrm{a}$ Likely  blended with a stellar line.

\end{table*}

For the S1 star, which is a member of the open cluster Danks 2 
\citep{davies_12}, the strength of the interstellar absorption 
features in its spectrum is very similar to that in the spectra of
\WC and \WR (see Table~\ref{tab:dibs}). This is a solid indication
that all the three objects are located at about the same distance from
us. Also, the interstellar extinction to S1 is consistent with that to 
\WC ($E_\mathrm{B-V}=2.73$ vs. $2.67$ mag; see
Section~\ref{subsubsec:wc8_fit}), which gives additional confidence 
for the results from our spectral fits.

The relatively small values of the interstellar extinction derived
for CPD-62~3058 and S2 indicate that
these objects are likely foreground stars. We note that the 
$E_\mathrm{B-V}$ value for CPD-62~3058 is in acceptable correspondence 
with that ($E_\mathrm{B-V} = 1.01$~mag) reported by \citet{mcgrud_75}. 
The situation with the
stars S3 and S4 is inconclusive because of the quality of their SALT
spectra (see also Table~\ref{tab:add_dibs}).

\section{Global spectral fits to the \WR optical spectra}
\label{append:fits}

The adopted technique for the global spectral fits to the optical
spectra of \WR was discussed in Sections~\ref{subsec:fits} and
\ref{subsubsec:wr48a_fit}. We only recall that a two-component fit  
was used in this analysis. Since \WR was classified as Wolf-Rayet star
of the WC8 type, the first component of the fit was the SALT
spectrum of the WC8 star \WC whose spectrum was derived simultaneously
with that of \WR (see Section~\ref{sec:observations} and
Fig.~\ref{fig:salt_slit}). For the 
second component in the \WR fit, we used spectra from STELIB library 
\citep{leborgne_03} or from the archives of various observatories.
The SALT and AAT spectra were fitted as the interstellar extinction
was also taken into account. Another important parameter derived in the
fits is the relative brightness of the stellar components. This is 
denoted  by the difference of the intrinsic V-magnitude of 
the WC8 star ($V_{WC8}$) and its companion star ($V_C$): 
$\Delta V = V_{C} - V_{WC8}$.

\begin{figure*}
\begin{center}
\includegraphics[width=3.in, height=2.15in]{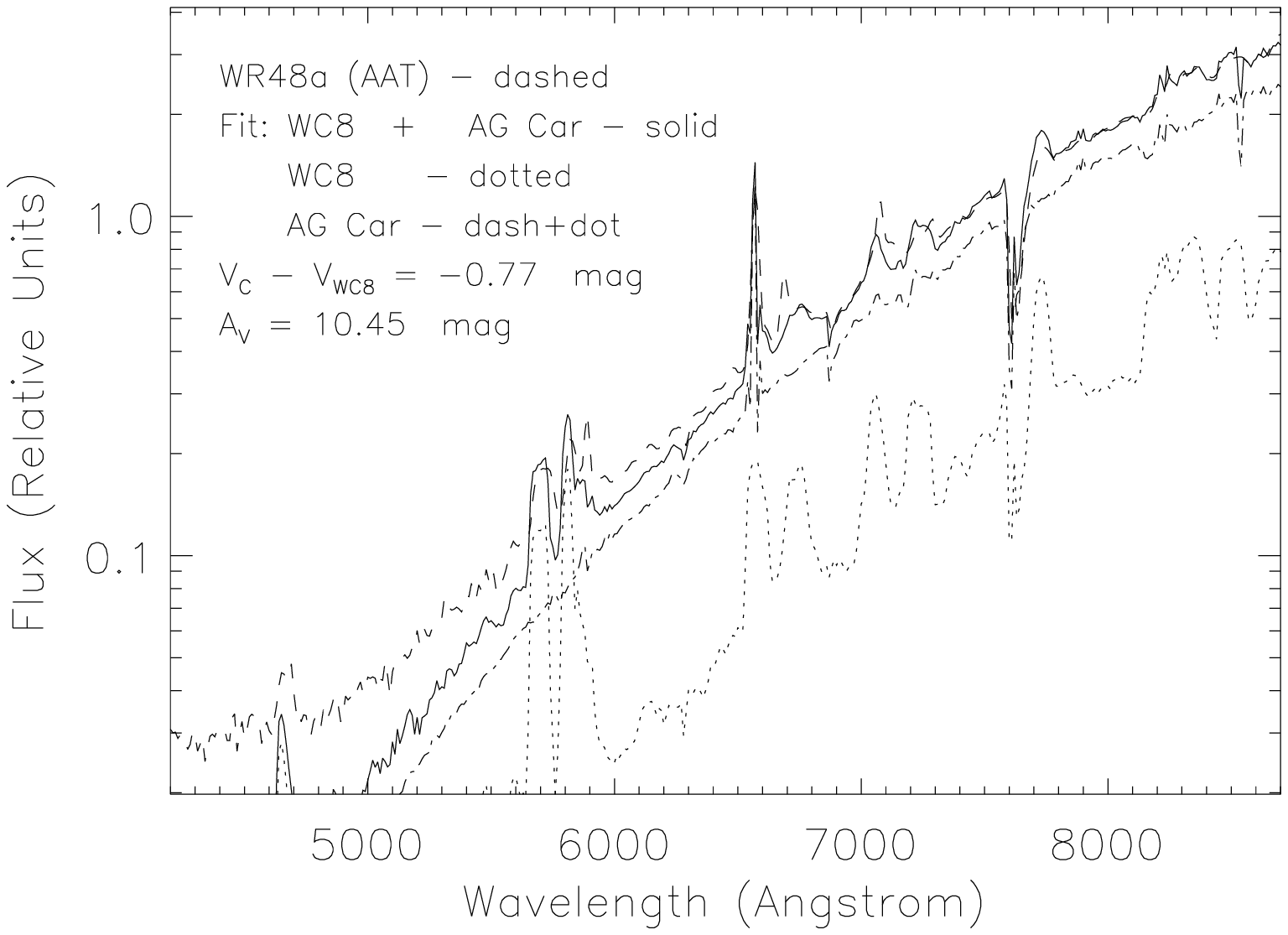}
\includegraphics[width=3.in, height=2.15in]{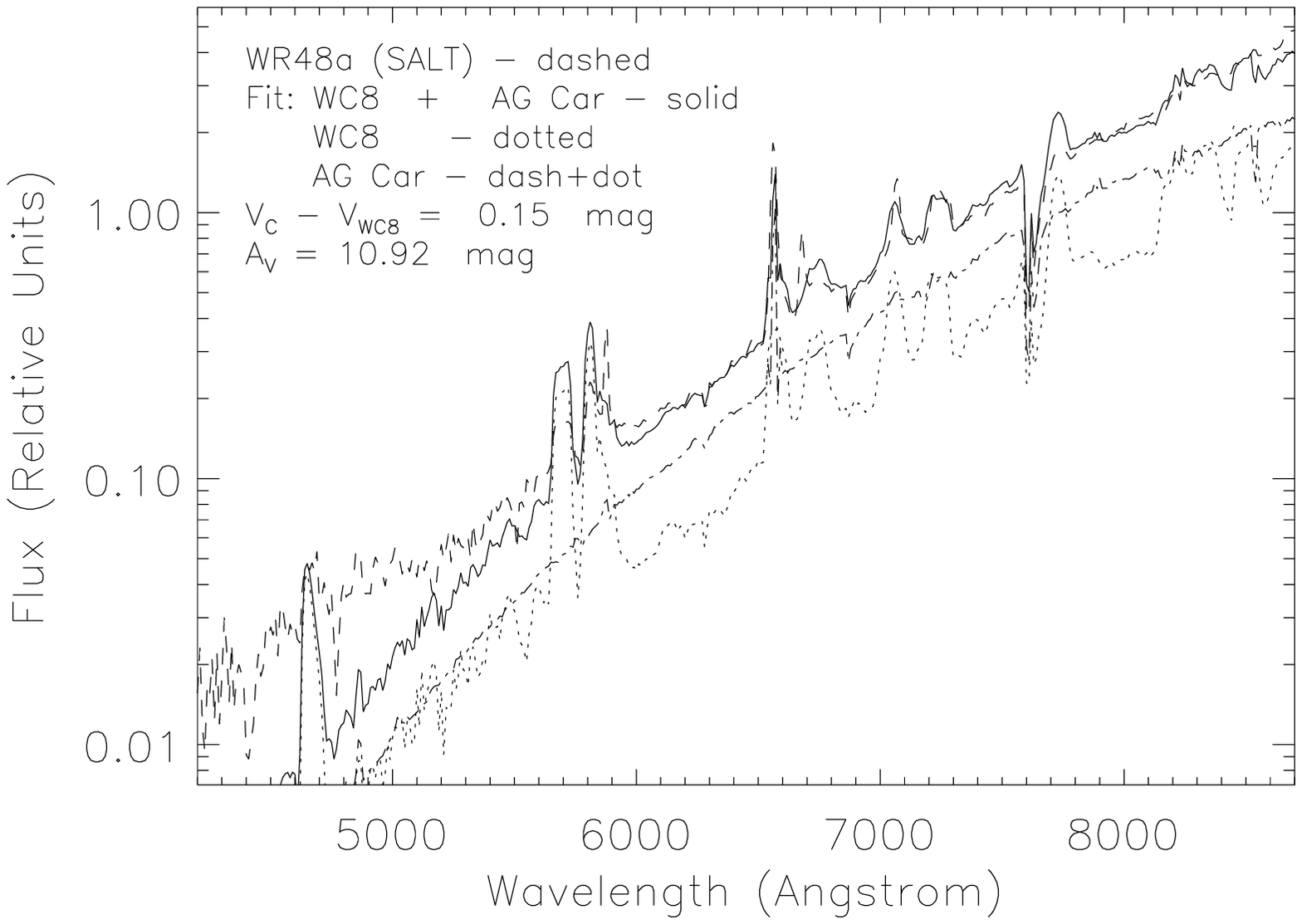}
\includegraphics[width=3.in, height=2.15in]{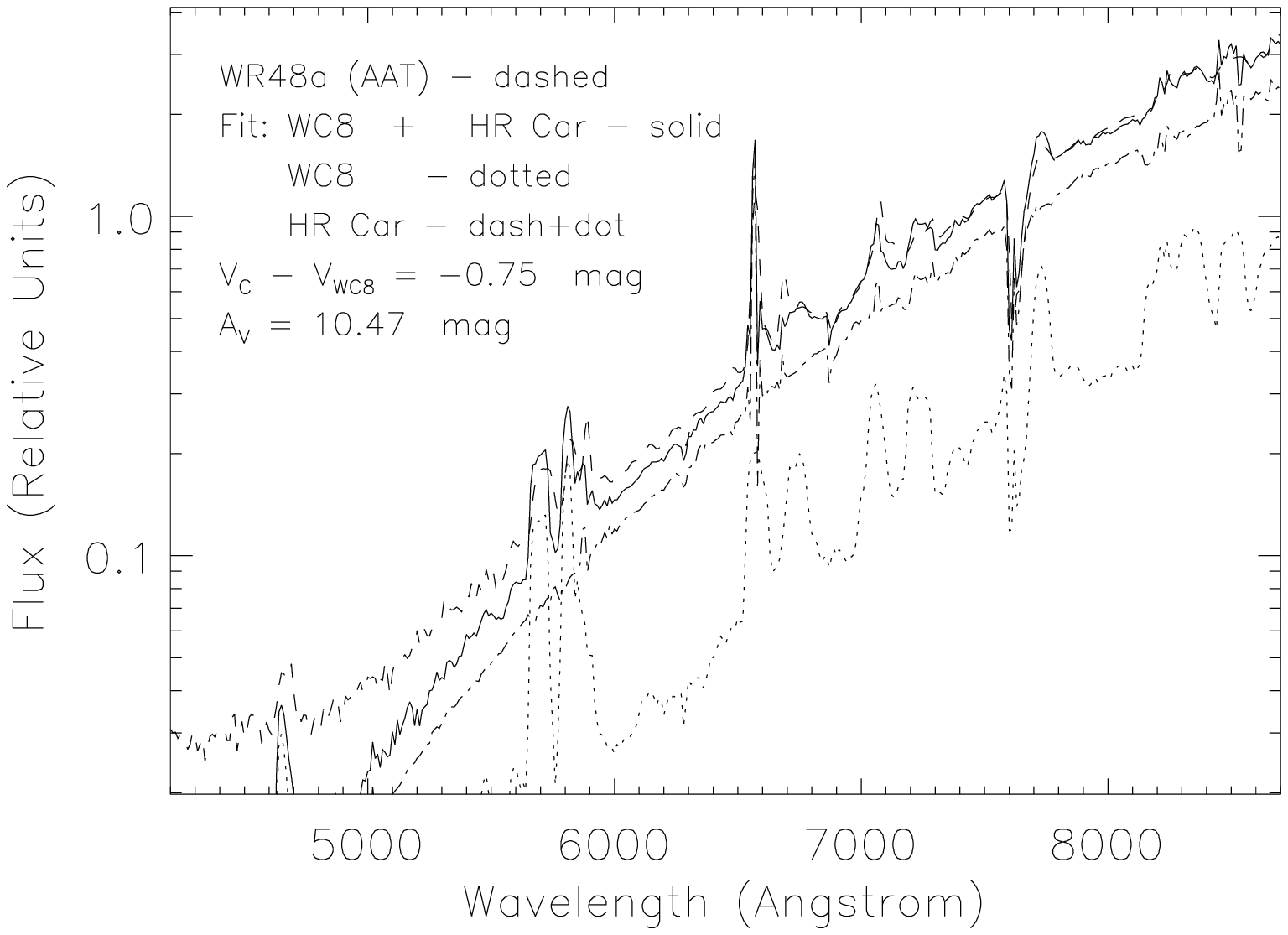}
\includegraphics[width=3.in, height=2.15in]{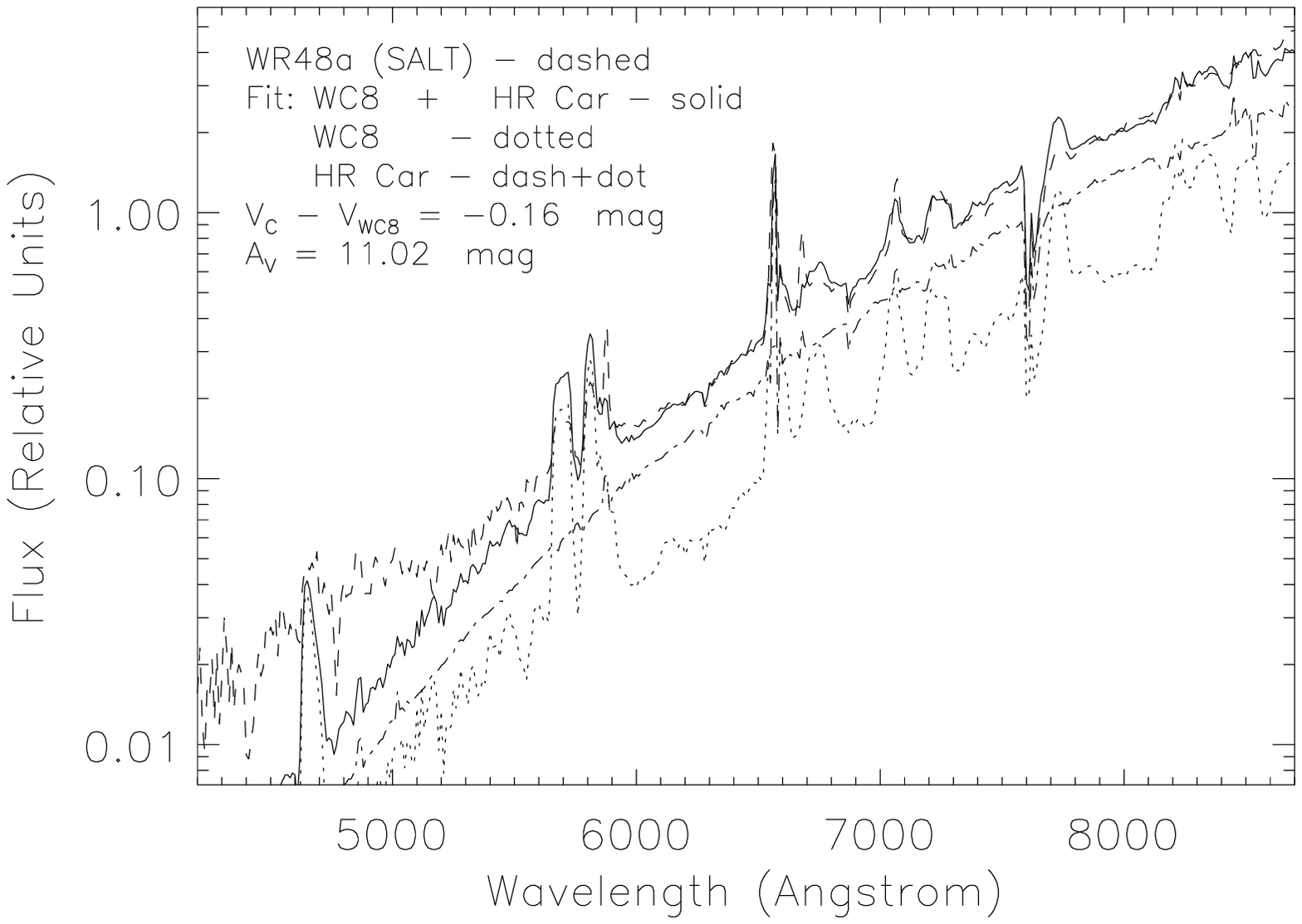}
\includegraphics[width=3.in, height=2.15in]{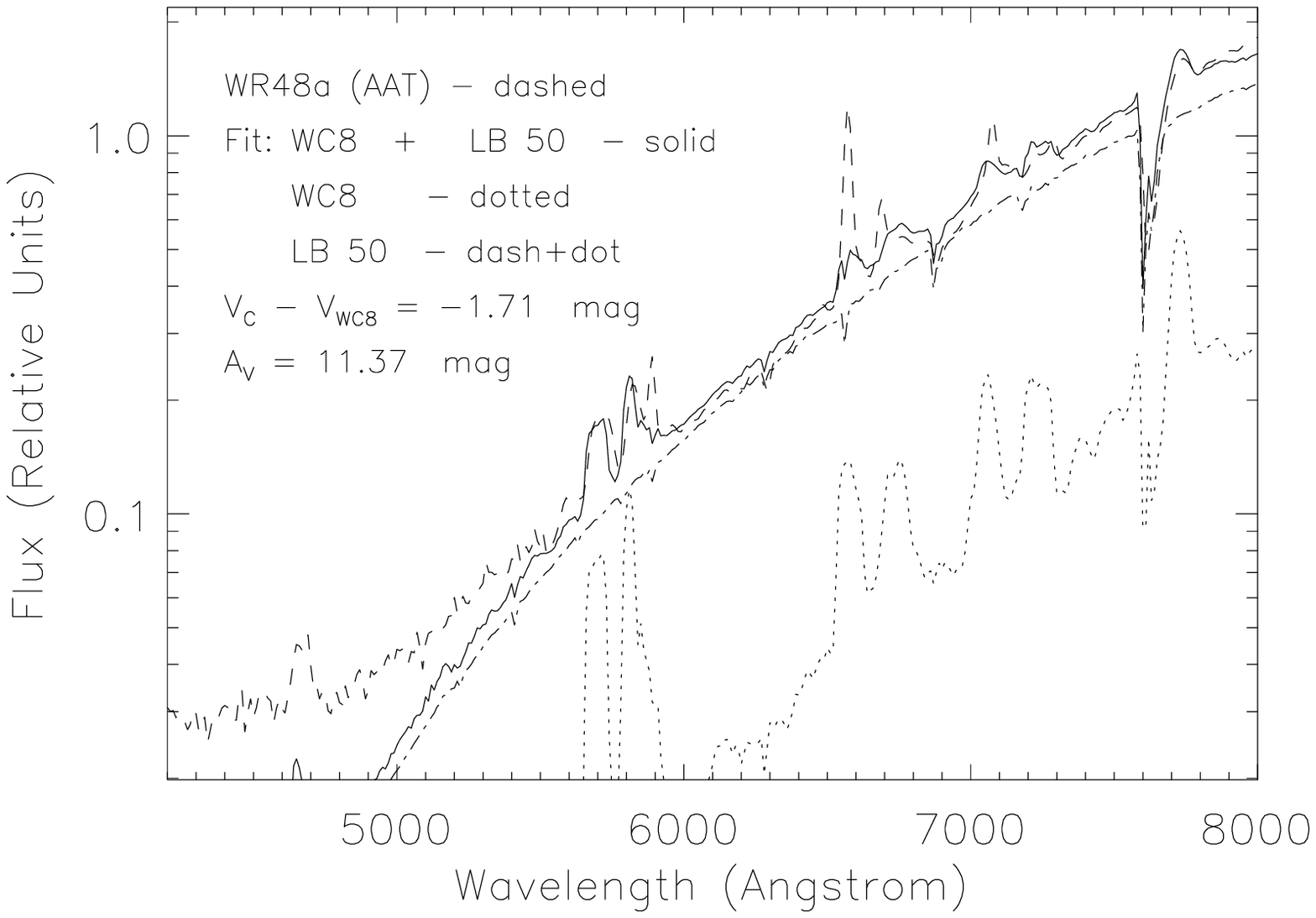}
\includegraphics[width=3.in, height=2.15in]{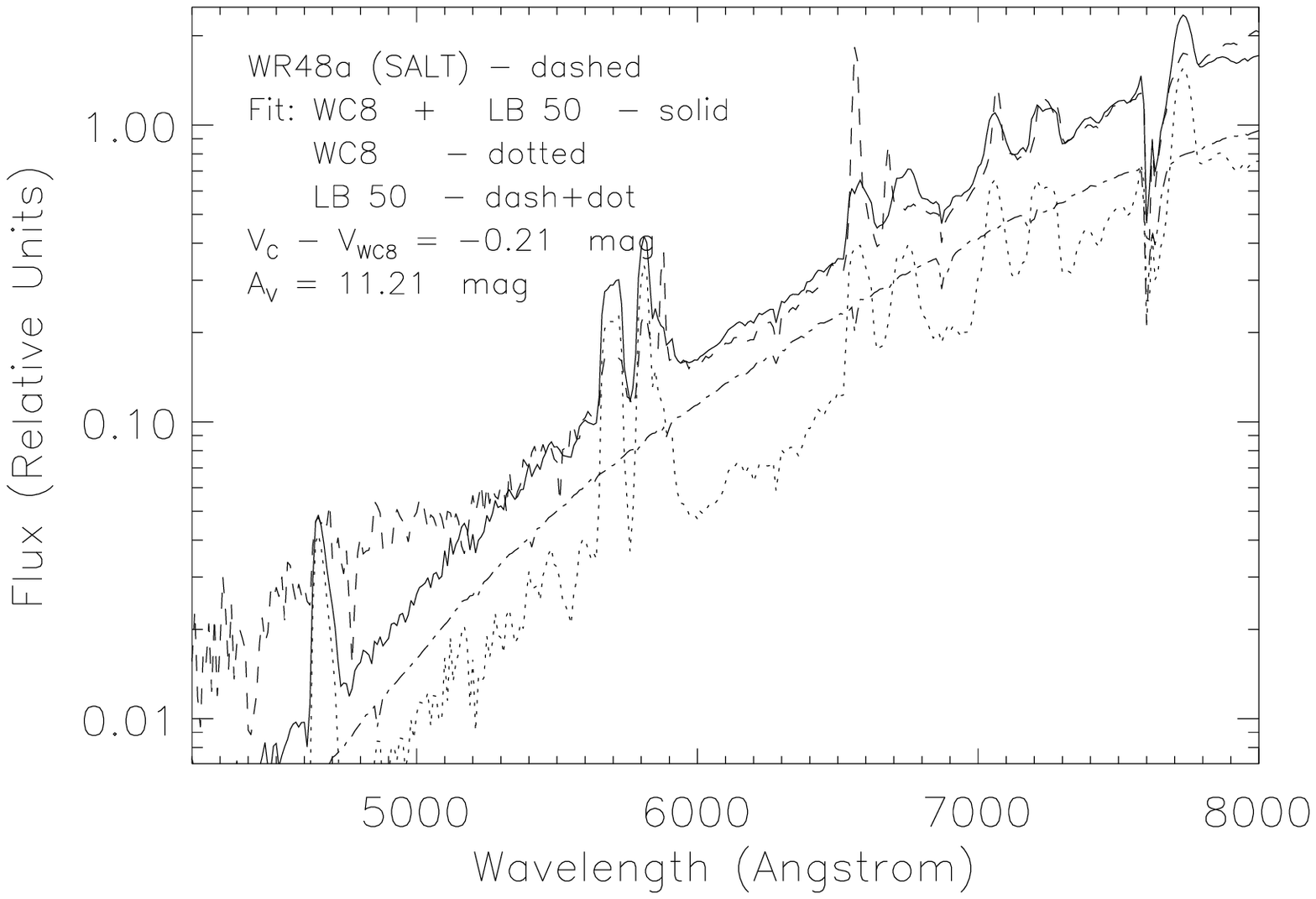}
\includegraphics[width=3.in, height=2.15in]{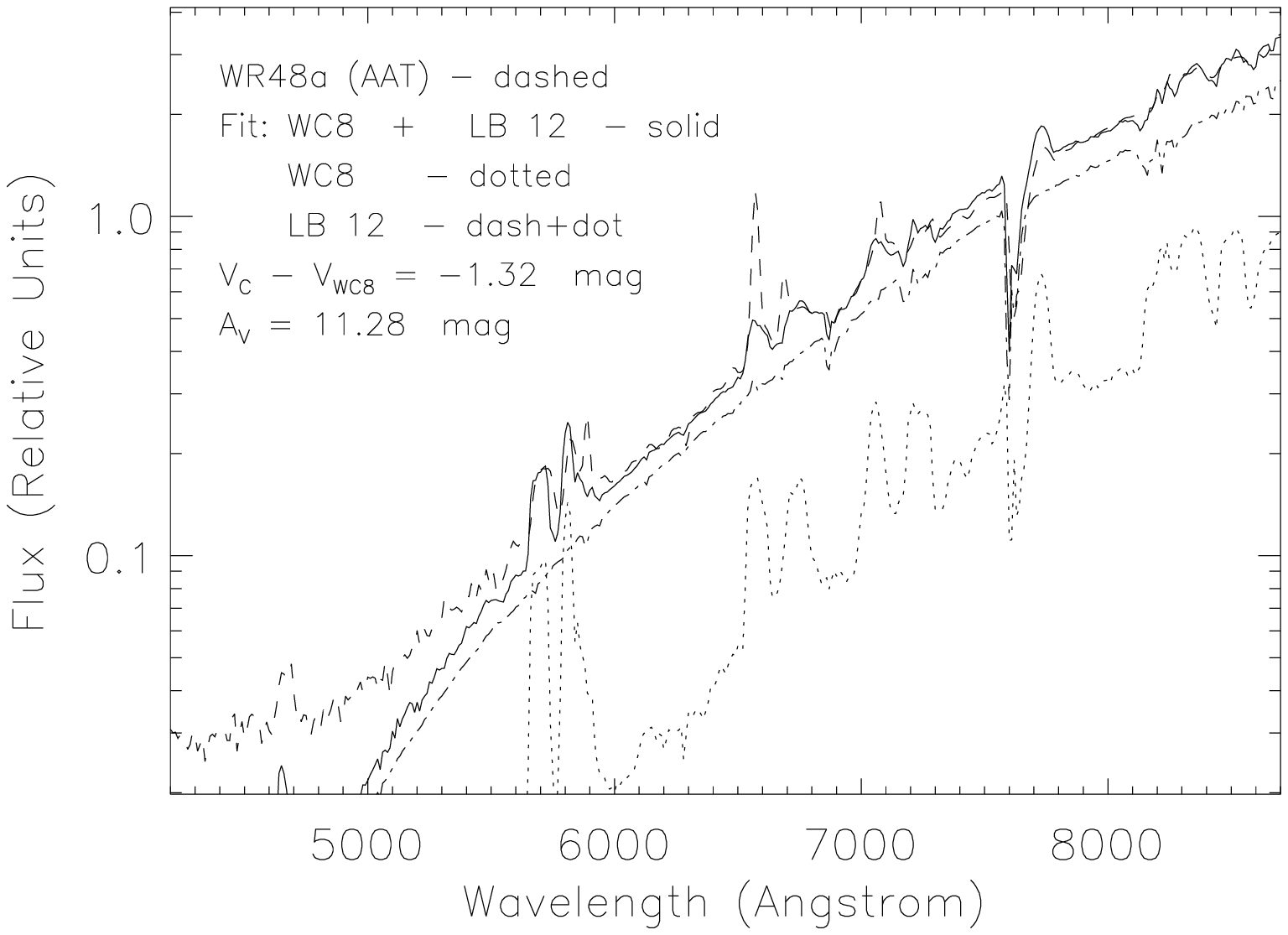}
\includegraphics[width=3.in, height=2.15in]{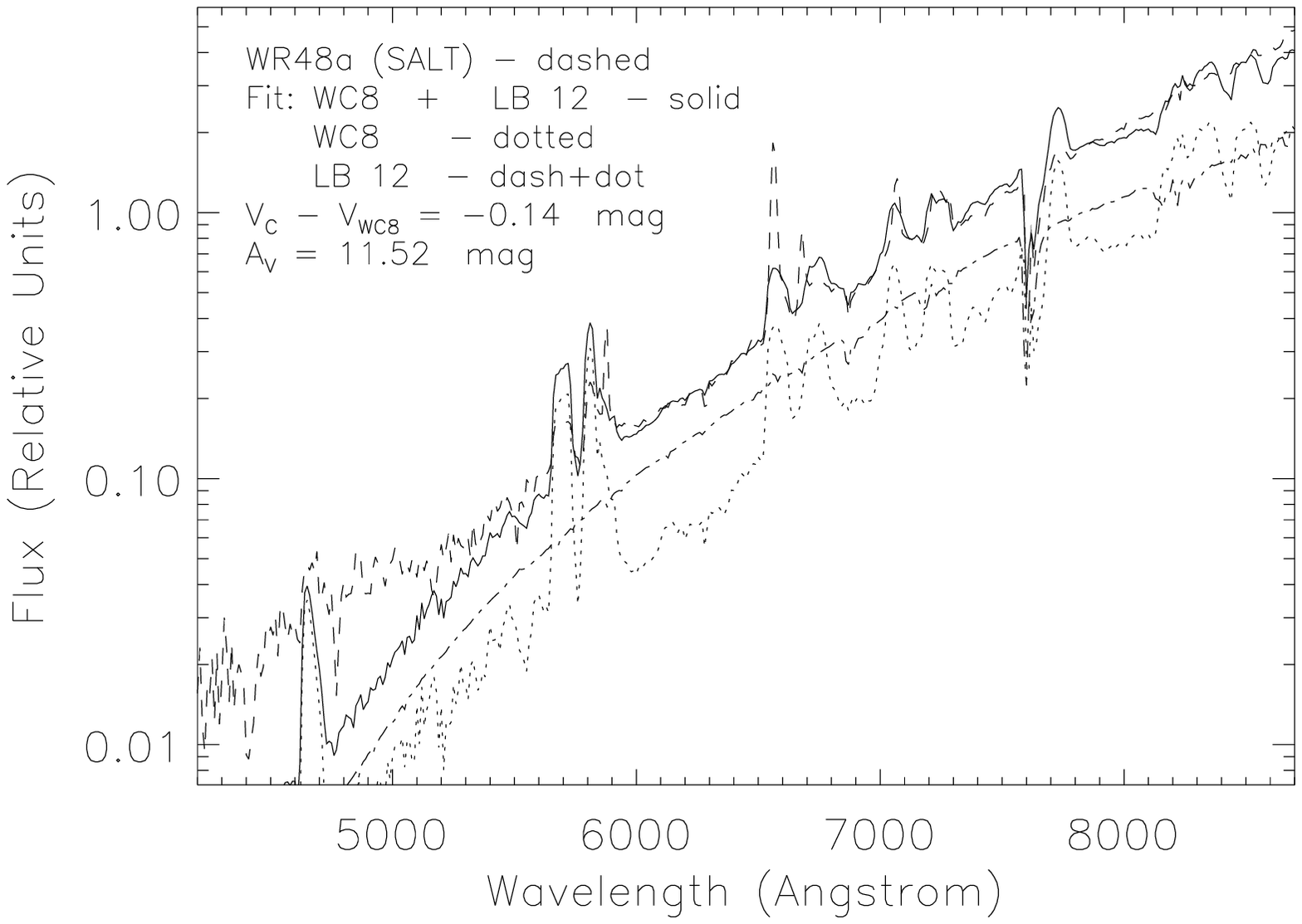}
\end{center}
\caption{Examples of the
two-component fits to the AAT and SALT spectra of \WRE. The first
component is the SALT spectrum of the \WC star and the second component 
is the spectrum of a LBV star (AG Car, HR Car), an O star (O5e; LB 50) 
and a B star (B2Ia; LB 12). The `LB' denotes that the data are from 
the STELIB library of stellar spectra \citep{leborgne_03}. The two 
digits represent the object number in the library. 
The bin size is 10~\AA.
}
\label{fig:lbv}
 \end{figure*}

The results from these global fits were presented in
Section~\ref{subsubsec:wr48a_fit} in the case of a WN8h spectrum as a
second spectral component. These fits gave the best match to the
optical SALT and AAT spectra of \WRE. Figure~\ref{fig:lbv}
%\ref{fig:o-stars} and \ref{fig:ab-stars} 
presents some results from the
global fits if the second component is a spectrum of a LBV, O or B
%A and B 
star, respectively. We see that these fits are not able to match 
well the relatively narrow spectral lines of helium and hydrogen 
(He\,{\sc i} 5876, 6678~\AA, H$_{\alpha}$~6563~\AA). On the other hand, these 
fit results confirm one of the basic findings discussed in
Section~\ref{subsubsec:wr48a_fit}. Namely, the interstellar extinction
in the SALT spectrum (2012 May) of \WR is higher than in its AAT
spectrum (1993 June). They also establish the relative variability
(change) between the strengths of the WC8 spectral component and that
of the companion star.

\bsp

\label{lastpage}

\end{document}